\documentclass[12pt,manuscript]{aastex}
\usepackage{graphicx}
\usepackage{epstopdf}
\usepackage{natbib}
\usepackage{amsmath}
\providecommand{\e}[1]{\ensuremath{\times 10^{#1}}}
\usepackage{url}

\shortauthors{Davenport et al.}
\shorttitle{The Destruction of M67}

\begin{document}

\title{Death of a Cluster: the Destruction of M67 as seen by the SDSS}

\author{James R. A. Davenport\altaffilmark{1,2},
Eric L. Sandquist\altaffilmark{3}}

\altaffiltext{1}{Corresponding author: jrad@astro.washington.edu}
\altaffiltext{2}{Department of Astronomy, University of Washington, Box 351580, Seattle, WA 98195}
\altaffiltext{3}{Department of Astronomy, San Diego State University, 5500 Campanile Drive, San Diego, CA, 92182-1221}

\begin{abstract}
We probe the spatial and dynamical structure of the old open cluster M67 using photometric data from the Sloan Digital Sky Survey's sixth data release. Making use of an optimal contrast, or matched filter, algorithm, we map the distribution of high probability members of M67. We find an extended and elongated halo of likely members to a radius of nearly $60'$. Our measured core radius of $R_{core} = 8\farcm24\pm0\farcm60$ is somewhat larger than that of previous estimates. We attribute the larger core radius measurement to the SDSS probing lower mass main sequence stars than has been done before for similar studies of M67, and the exclusion of post main sequence M67 members in the SDSS sample. We estimate the number of M67 members in our SDSS sample to be $1385\pm67$ stars.  A lower limit on the binary fraction in M67 is measured to be 45\%. A higher fraction of binary stars is measured in the core as compared to the halo, and the luminosity function of the core is found to be more depleted of low-mass stars. Thus the halo is consistent with mass segregation within the cluster. The galactic orbit of M67 is calculated from recent proper motion and radial velocity determinations. The elongated halo is roughly aligned to the proper motion of the cluster. This appears to be a result of mass segregation due to the galactic tidal field. Our algorithm is run on 2MASS photometry to directly compare to previous studies of M67. Decreasing core radii are found for stars with greater masses.
We test the accuracy of our algorithm using 1000 artificial cluster Monte Carlo simulations. 
It is found that the matched filter technique is suitable for recovering low-density spatial structures, as well as measuring the binary fraction of the cluster.
\end{abstract}

\keywords{open clusters and associations: individual (M67) -- galaxy: kinematics and dynamics}

\section{Introduction}
For over 40 years the spatial structure of stellar clusters has been characterized by the widely accepted King model \citep{king62}. This density model matches a wide range of observed clusters well 
 \citep[e.g.][]{bigcluster}. As \citet{king66} notes, however, the model does not describe the specific dynamical evolution or state of a particular cluster. As clusters age their constituent stars inevitably undergo close gravitational encounters with other members of the cluster. Through energy equipartition, lower mass stars are given higher velocities, and thus larger orbits in the cluster. Many will be given velocities greater than the escape speed for the cluster's gravitational well.  We  observe this dynamical equipartition in the segregation of masses radially across a cluster, with lower mass stars being found preferentially further from the center than higher mass stars. Binary stars are also more centrally concentrated than their single-body counterparts. The King model does not deal with the  
unique history a cluster may have within its parent galaxy, or the state and location of stars previously associated with it.

Escaped stars may be projected across large areas on the sky. Within disrupting clusters (such as most open clusters for example) the ejected stars may be a significant fraction of the cluster's initial mass. It is also likely that the ejected stars are not distributed in a spherical manner, due to tidal disruptions from the galactic potential or close encounters with giant molecular clouds. By understanding the total amount and rate of mass lost in a cluster, we may begin to paint a picture of the cluster initial mass function, and the history of its disruption in the galaxy. Since open clusters are plentiful and found with a wide variety of ages, we may hope to study this process at many stages.

M67 (NGC 2682) is an old ($\sim 4$Gyr), nearby ($\sim$900pc), and well-studied open cluster. Its proximity has made it an ideal target for a wide array of studies over the past century, ranging from radial velocities \citep{1986AJ.....92.1100M} and proper motions \citep{sanders77,1989AJ.....98..227G}, to late-type stellar evolution and binary fractions \citep{1993AJ....106..181M}.  

The disruption scenarios and timescales of open clusters in our galaxy have been studied for some time \citep[e.g.][]{spitzer57, spitzer58}. Recently N-body models have become sophisticated enough to place constraints on the role of giant molecular clouds and detailed initial structure in cluster dissolution \citep[e.g. see][]{gieles06, gieles08}.
Since most open clusters are thought to generally dissipate over $\sim10^7$ year timescales \citep{bt87}, the existence of old open clusters is somewhat of an anomaly.  While M67 is a classic example of a highly evolved open cluster \citep[e.g.][]{vdb57}, older examples do exist. NGC 6791 has an age of at least 8Gyr \citep{carraro06}, roughly twice that of M67. This open cluster is believed to be one of the most massive in our galaxy, and its high stellar density is surely a prerequisite for its survival to such exceptional age. Clearly this is an abnormal open cluster, and its origins under the typical open cluster formation scenarios have been questioned \citep{carraro06}. Despite its uncertain origins, NGC 6791 like other older clusters displays a dynamically evolved mass function, indicative of mass segregation \citep{1995A&AS..114....1K}.

Due to the paucity of old open clusters such at M67 and NGC 6791, and the large number of younger clusters, it is reasonable to assume that a great many open clusters existed in the past and have since been destroyed. These lower mass clusters, like the young clusters found today, likely had a wide range of masses and numbers of members. Indeed it is believed that most stars which are now part of the galactic disk originated in clusters of various sizes. These clusters must have had densities that were sufficiently low as to allow them to be completely dissociated within about 600 Myr \citep{bergond01}.

Ongoing work to fully model the evolution of M67 over its entire lifespan has been promising. \citet{m67nbody} have created an N-body simulation of M67 whose result over 4 Gyr of evolution shows reasonably good agreement with observations of the spatial distribution and stellar composition of the cluster. These types of simulations allow observers to study the cluster's initial conditions and their effect on the present state. M67 is found in this model to have lost at least 75\% of its stellar mass to dynamical and stellar evolution, and it is suggested that the cluster will continue to dissipate.

 Mass segregation has previously been suggested as the cause for the observed structural properties of M67. \citet[hereafter BB03]{bb03} find a notable difference between the luminosity functions of the core and halo of M67, with the halo having significantly more faint star contributions. This is found as well in several other studies, such as \citet{1993AJ....106..181M} and \citet{fan96}
 
BB03 make use of the Two Micron All Sky Survey \citep[hereafter 2MASS]{2mass} to probe M67 over a large and continuous area. M67 has a large enough spatial projection on the sky that it could not be imaged by most large telescopes in a single exposure. Unlike dedicated ``pencil-beam'' surveys which could be carried out over individual portions of the cluster, large scale surveys such as 2MASS and the Sloan Digital Sky Survey \citep[hereafter SDSS]{york2000} can uniformly image regions much larger than a single cluster. This makes them ideal resources for studies of cluster structure.
 
In this paper we use SDSS data to probe the spatial and stellar structure of M67 as it is seen today. Our selection of SDSS photometry is described in \S2. The matched filter technique which separates M67 from the surrounding field population is outlined in \S3, and our results are presented in \S4. In \S5 we offer a discussion of the results and a calculation of the orbit of M67, as well as a summary to put our findings in context with previous work on M67.

\section{SDSS Photometry}
The SDSS is a major survey consisting of both five-band ({\it ugriz}) photometric and optical spectroscopic data. The photometric data are collected in adjoining stripes over a quarter of the sky, covering the north galactic cap. 
It has been used to study objects ranging from nearby asteroid families \citep{2008Icar..198..138P} to incredibly distant quasars \citep{2009AJ....137.4118I}. The SDSS provides nearly simultaneous imaging in five photometric bands ({\it ugriz}), down to r$\sim$23 \citep{dr6}. These data are carefully calibrated \citep{tucker06, davenport07}, achieving excellent photometric errors averaging 1\% in {\it u,g,r,i} and 2\% in {\it z}. The large spatial coverage combined with accurate photometry has been useful for studying low-density structures such as tidal streams in the galactic halo and globular cluster disruptions \citep[e.g.][]{2006ApJ...643L..17G}. The SDSS thus provides an ideal source of information about several nearby open clusters, which are often distributed over large portions of the sky.

Our data came from the sixth SDSS data release \citep[hereafter DR6]{dr6} which is publicly available online via a SQL database\footnote{\url{http://casjobs.sdss.org}}.  The reader is referred to DR6 and references therein for a highly detailed description of the survey. Our query returned sources from the STARS view of the PhotoObjAll table, which includes all objects which the SDSS pipeline \citep{lupton01} detects as point sources (as opposed to extended sources or cosmic rays for example) and are not quasars. To ensure a clean sample of only stars in our analysis we  set the following photometric flags: SATURATED = 0, BRIGHT = 0, OK\_SCANLINE $>$ 0, OK\_STRIPE $>$ 0, EDGE = 0, LOCAL\_EDGE = 0, PRIMARY $>$ 0, PSF\_FLUX\_INTERP = 0, INTERP\_CENTER = 0, BAD\_COUNTS\_ERROR = 0, PEAKCENTER = 0, NOTCHECKED = 0. Some of these flags are redundant to those automatically applied when querying from the STARS view. The stripes of the survey run across great-circles along galactic latitude. We found flags such as OK\_SCANLINE, OK\_STRIPE, and PRIMARY particularly necessary to remove over-densities of stars found within the overlaps of the stripes. The SDSS provides a description of suggested use of flags online at \url{http://www.sdss.org/dr6/products/catalogs/flags.html}.  We have chosen to use the PSF magnitudes as they are considered the most unbiased measurement for point sources by the SDSS.\footnote{\url{http://www.sdss.org/DR6/algorithms/photometry.html}}

\begin{figure}[]
\centering
\includegraphics[width=5in]{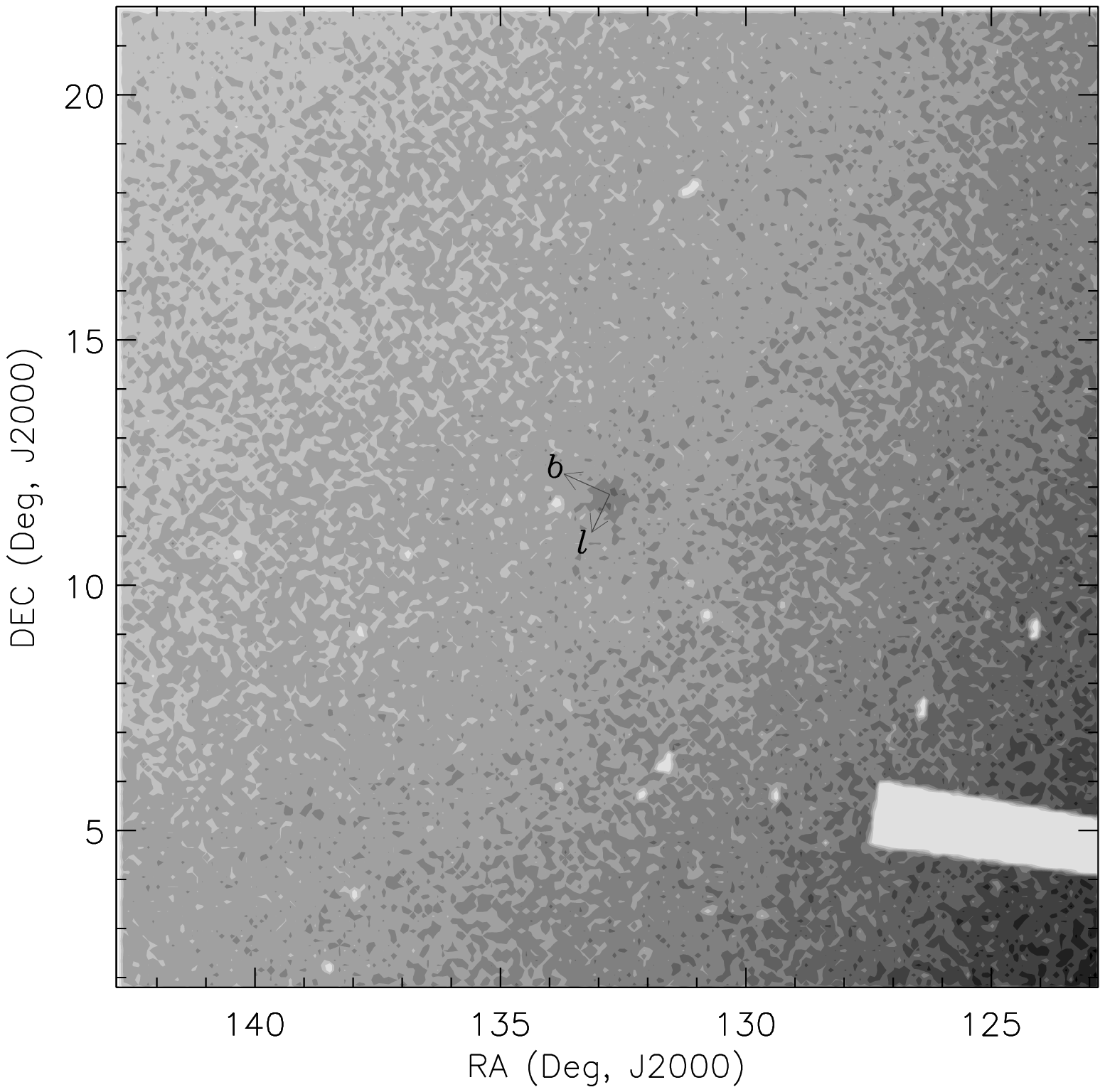}
\caption{The $20^\circ\times20^\circ$ SDSS DR6 field around M67. Darker contours indicate a higher density of stars. The stellar density ranges from roughly 3300 to 11100 per deg$^2$, with the density near M67 at $\sim$5000 per deg$^2$. Contours are in increments of 2000 stars per deg$^2$. M67 is faintly visible at the center of the field, and increasing galactic coordinates are marked from its center. The swath of missing data starting at about $(\alpha=123,\delta=5)$ is due to a missing piece of one stripe of the survey. Roughly a dozen white spots are seen, which mark locations of bright stars or galaxies which were masked out of the point-source catalog, including a bright star near M67.}
\label{field}
\end{figure}

We queried a $20^\circ \times 20^\circ$ box centered around M67 ($\alpha$=132.825, $\delta$=+11.8, J2000) which returned 2,111,236 point sources with $15<g<23$ and $14<r<22.5$. These magnitude cuts were made to remove major photometric scatter which grows rapidly near the faint limit, and to avoid bright stars which may have spurious photometry. The entire sample is shown in Figure \ref{field}, which shows the density of point sources found by SDSS as filled contours (where darker is increasing density).  Estimated photometric reddening is computed by the SDSS from the \citet{schlegel98} dust maps. We carried out the analysis in the following sections both with raw magnitudes and reddening corrected magnitudes. These reddening adjustments are critical for extragalactic or distant cluster science as they correct for the extinction of light from the interstellar medium. Because M67 is quite close, extinction is not suspected to be a major contribution to the observed photometry. Indeed, no significant differences arose in our results when using de-reddened magnitudes. The final analysis was completed with the de-reddened data.

\section{The Matched Filter}
The matched filter technique, also known as optimal contrast filtering, has been used in galactic astronomy for some time.  \citet[hereafter R02]{r02} used a matched filter on early SDSS photometry around the globular cluster Palomar 5. They were able to detect large symmetric tails of stars leading and trailing the cluster in its orbit. Follow-up matched filter work by \citet{2003AJ....126.2385O} and \cite{22deg} showed these tails to extend at least 22$^\circ$, a limit imposed by the edge of the SDSS footprint. We used the DR6 data around Palomar 5 as a benchmark for our algorithm, and were able to recover the $\sim$22$^\circ$ tails. A similar algorithm has been employed to detect low-density streams of stars with no detectable nucleus \citep[e.g.][]{2006ApJ...639L..17G,2006ApJ...643L..17G}. R02 provide a good description of the matched filter process, and the reader is encouraged to seek further discussion therein. We will however briefly describe our method and its application to the open cluster M67 as well as our artificial clusters below.

\subsection{Analysis of M67 from SDSS}
Traditionally in the absence of more detailed information, such as proper motions, radial velocities, or spectra, cluster membership can be crudely estimated based on the positions of stars in a color vs magnitude diagram (CMD). To maximize the contrast of the cluster against the surrounding field stars we must characterize the probability functions from both the field and cluster in color-magnitude space. These two CMD probability functions are then used to determine the probability of membership for every star.

\begin{figure}[]
\centering
\includegraphics[width=3in]{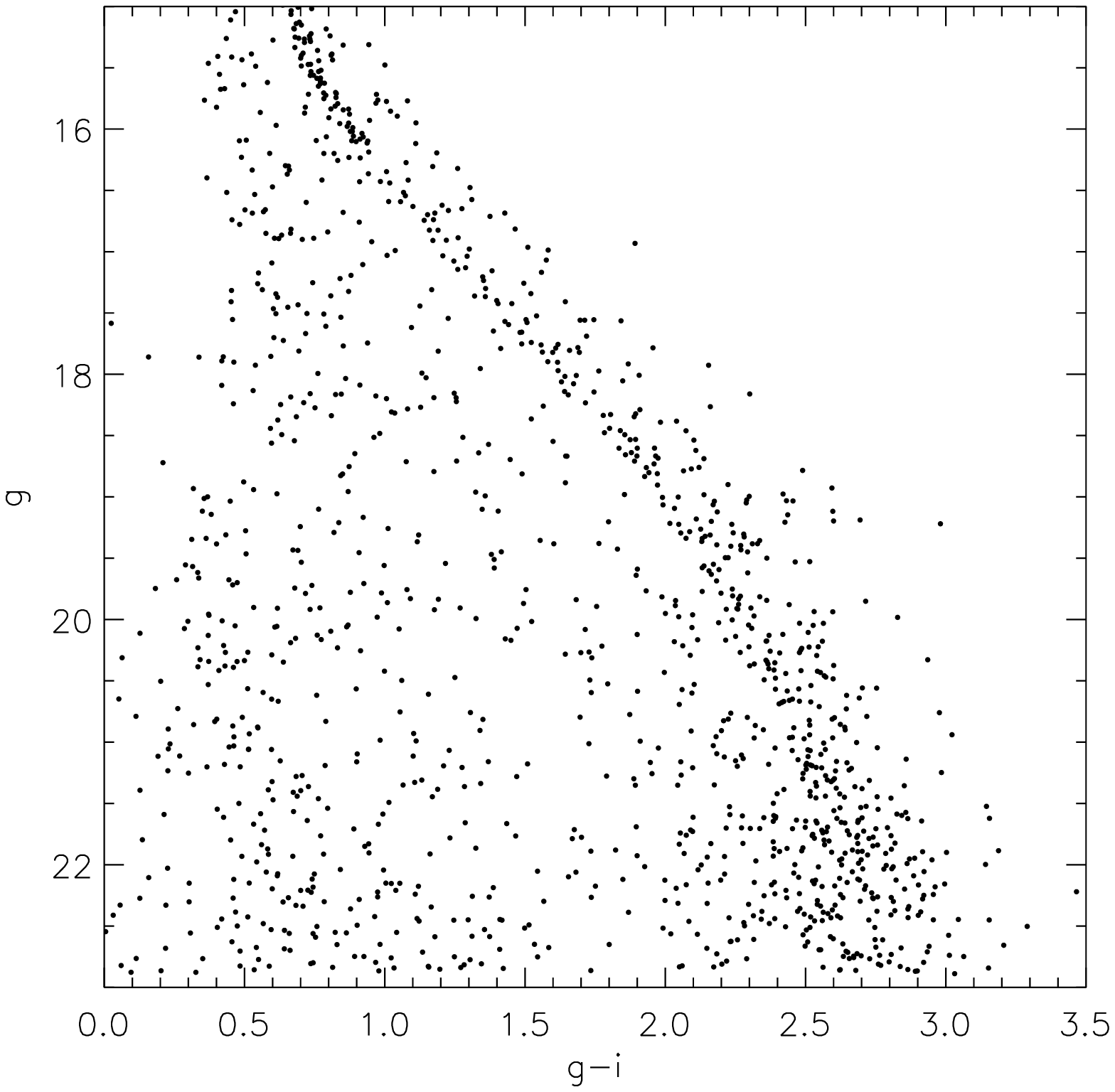}
\includegraphics[width=3in]{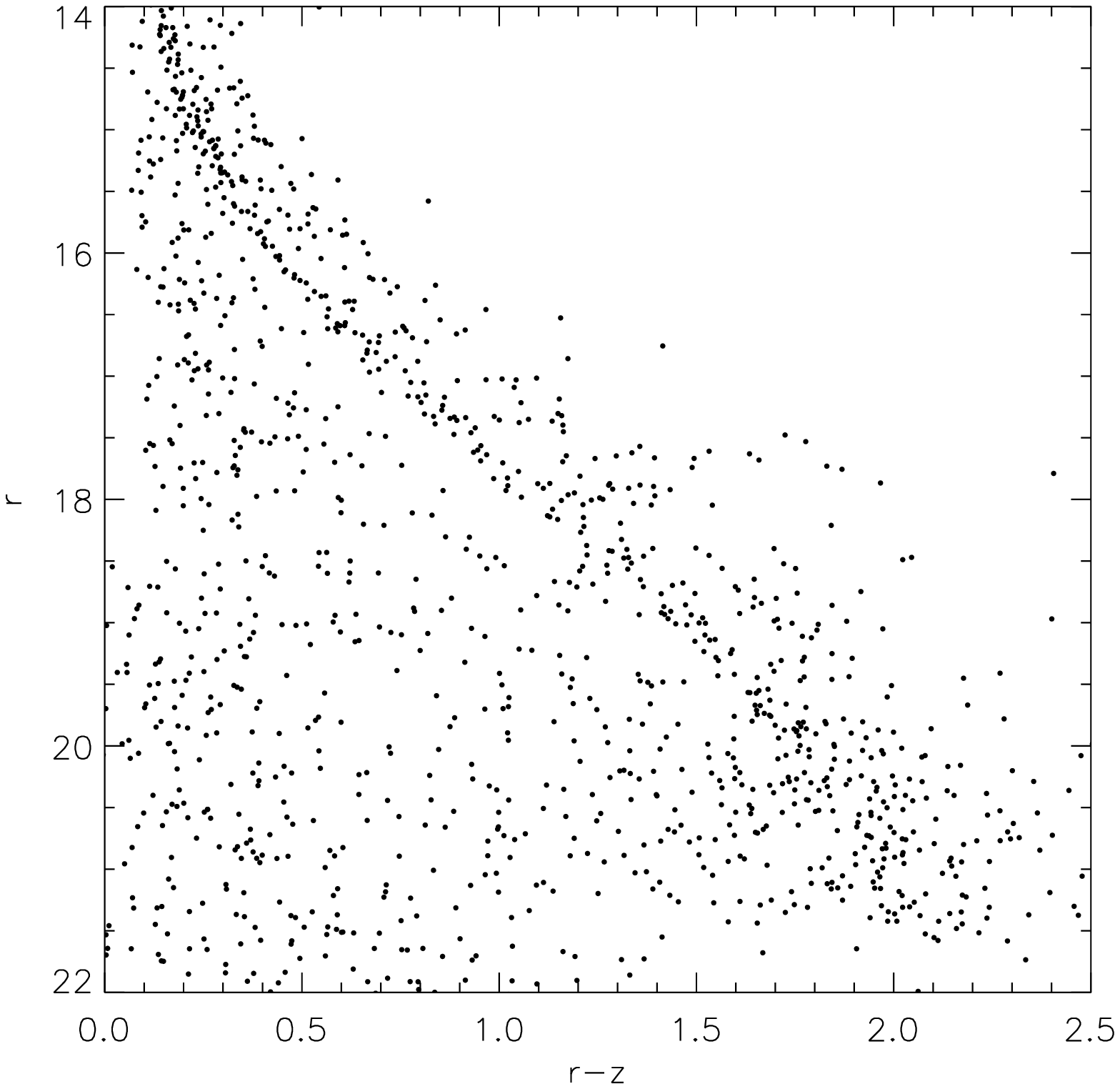}
\caption{The $(g-i,g)$ and $(r-z,r)$ CMDs of all 1389 stars within a radius of 0.25$^\circ$ from the center of M67 in our data.}
\label{cmds}
\end{figure}

The M67 CMD was determined by selecting the stars within $0.25^\circ$ of the cluster center. Figure \ref{cmds} shows the M67 CMDs for $g-i$ vs $g$ and $r-z$ vs $r$ in our SDSS data. The main sequence is clearly visible in both panels, and a faint equal-mass binary sequence can be seen above it. We chose to use the colors $g-i$ and $r-z$ because they better separate the M67 main sequence from the significant field contamination. The galactic field stars were crudely filtered out by rejecting any stars not within a few photometric $\sigma$ of a spline hand-fit to the main sequence and equal mass binary sequence.  A 2-D gaussian smoothing algorithm was  then applied to these rough cluster CMDs, creating a continuous CMD distribution for M67. 

\begin{figure}[]
\centering
\includegraphics[width=3in]{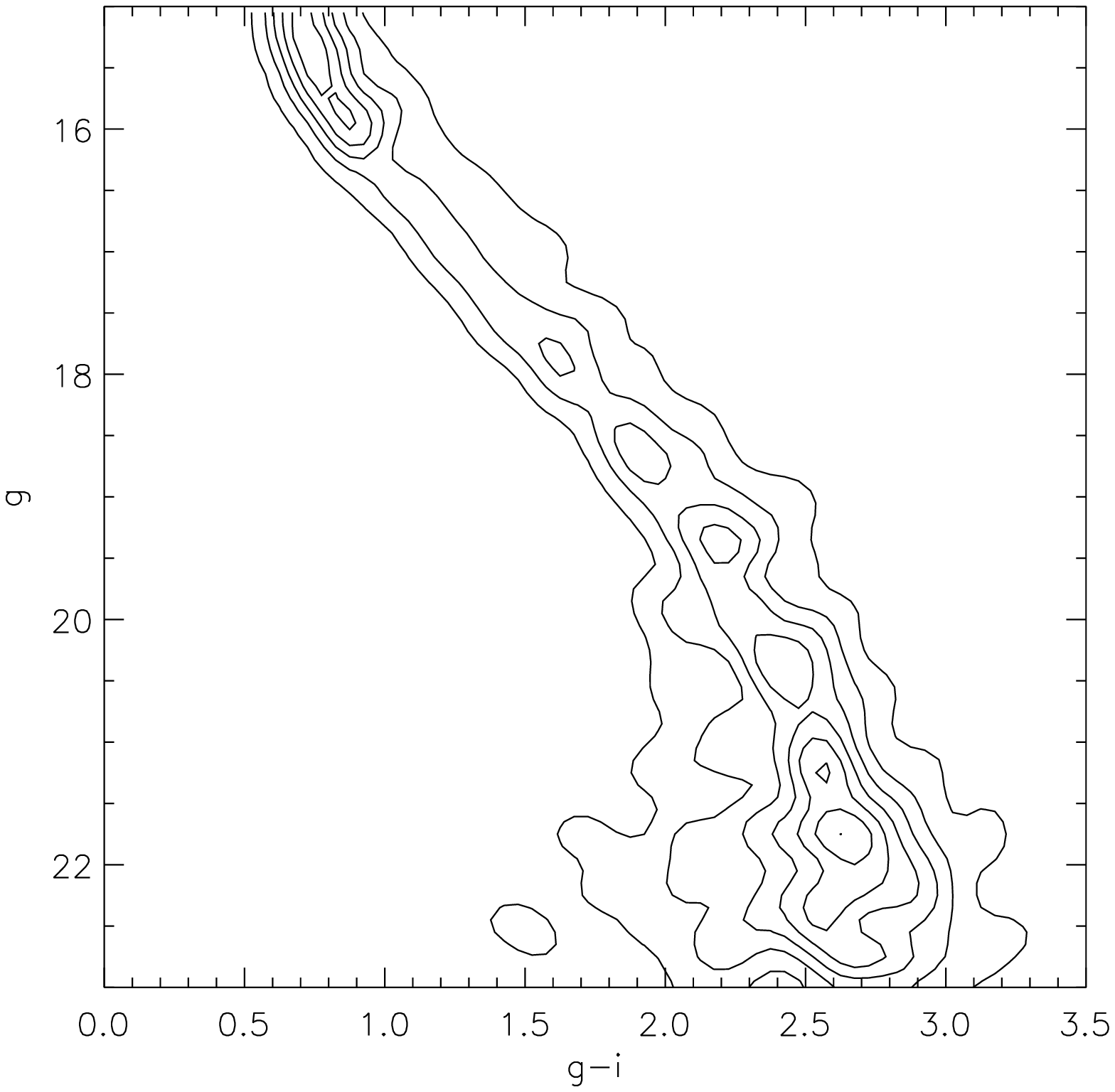}
\includegraphics[width=3in]{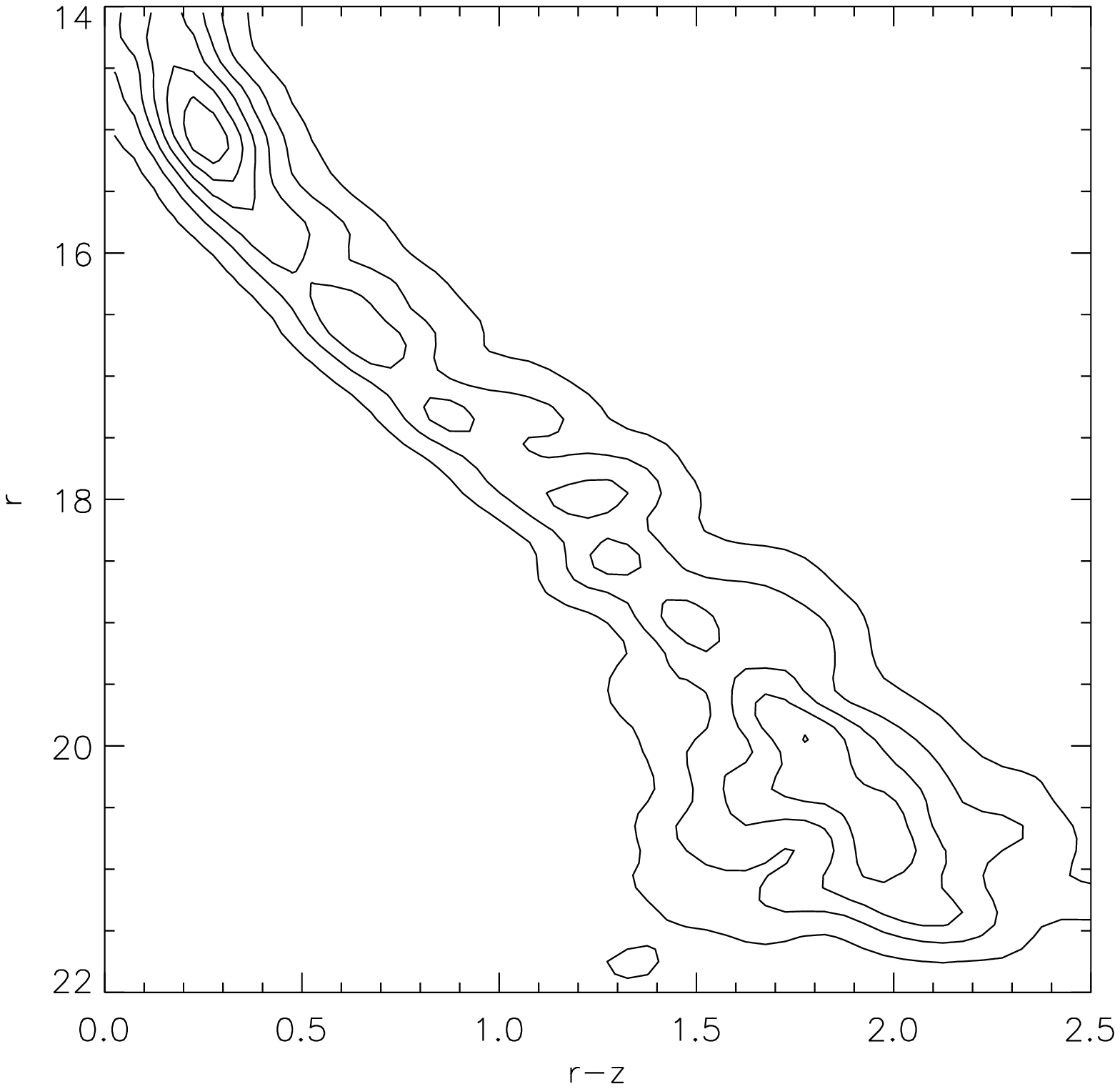}
\caption{The normalized and smoothed CMD filters M67, $f_{cl}(g-i,g)$ (left), and $f_{cl}(r-z,r)$ (right). The contours are steps of 1$\sigma$ above the mean level of the filter.}
\label{fcl}
\end{figure}

The number of cluster stars in a given solid angle is described by the equation $n_{cl} = \alpha f_{cl}$, where $\alpha$ is the number of stars in a region, and $f_{cl}(color,mag)$ is a normalized probability function for the cluster in the color-magnitude plane. By normalizing the continuous CMD distributions described above, we created the $f_{cl}$ functions for M67.
This was done for both $(g-i,g)$ and $(r-z,r)$ independently, and are shown in Figure \ref{fcl}. The non-uniform distribution shown in Figure \ref{fcl} is a result of sampling $f_{cl}$ from the stars in the core of M67 which are not evenly distributed along the main sequence. R02 investigated whether the mass function of the cluster core would detract from the matched filter's ability to detect tidal debris that would generally have a different mass function. Their conclusion is that the method is robust against such biases, but we do bear this in mind in our analysis later on.

\begin{figure}[]
\centering
\includegraphics[width=3in]{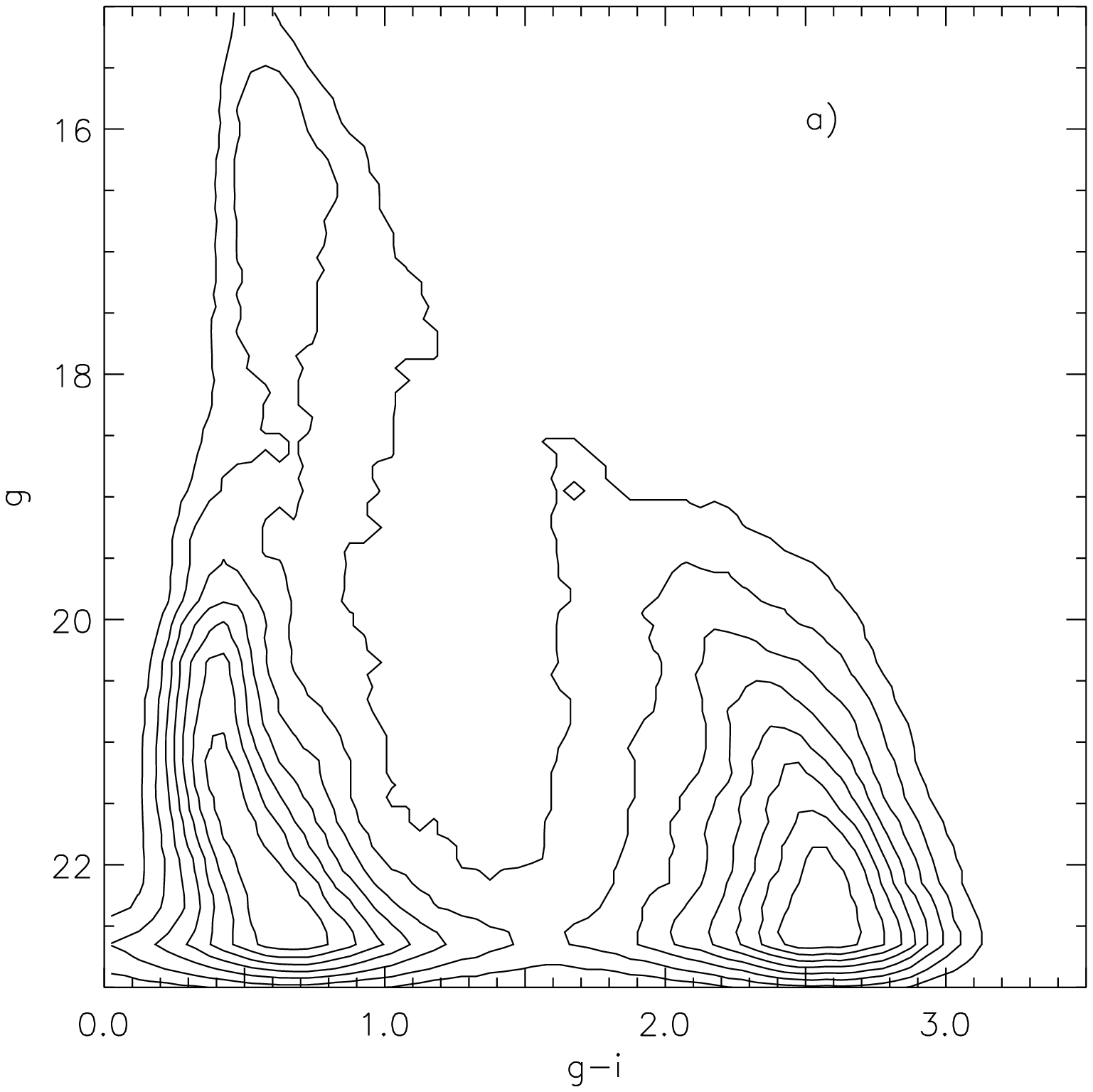}
\includegraphics[width=3in]{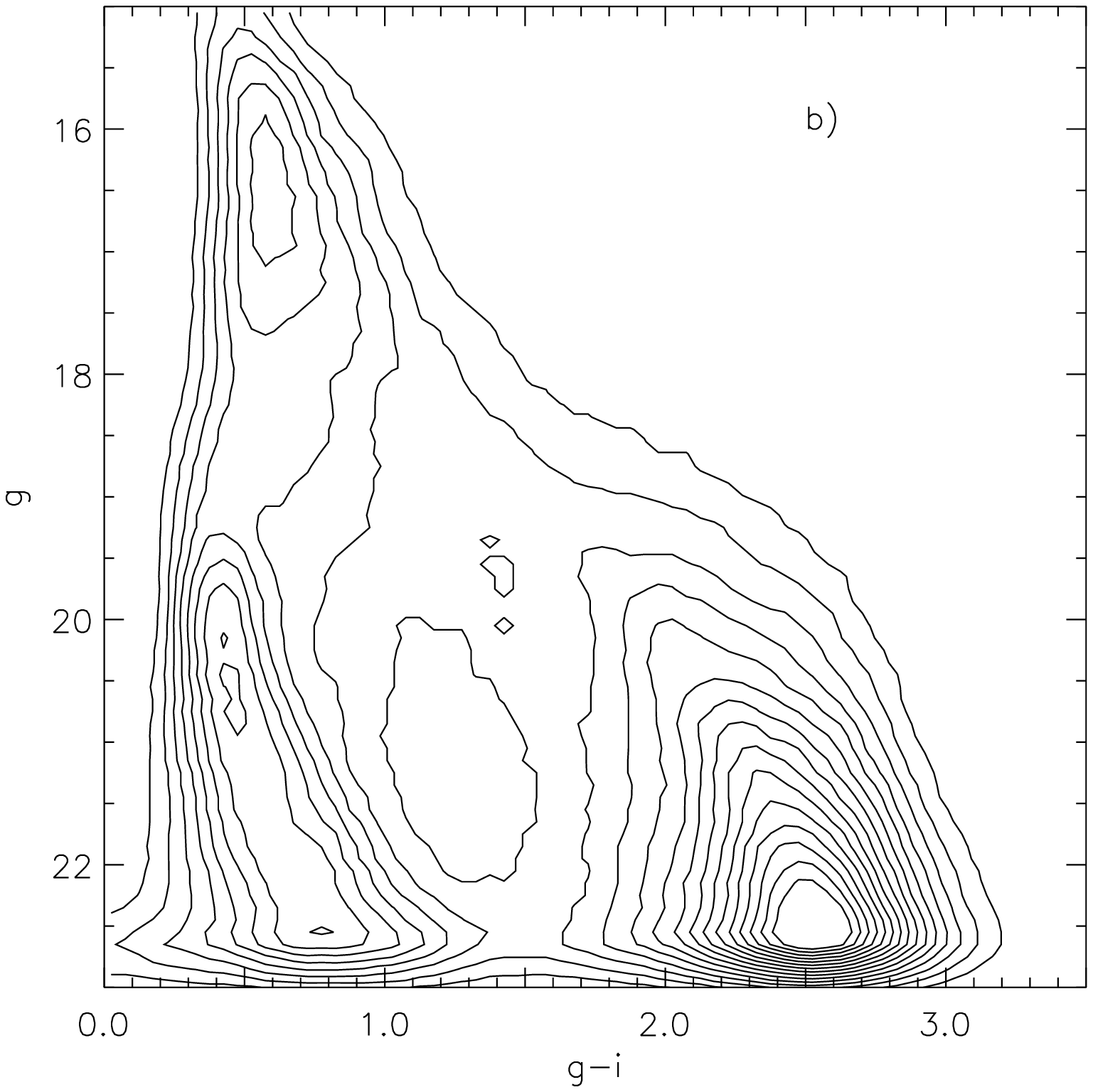}\\
\includegraphics[width=3in]{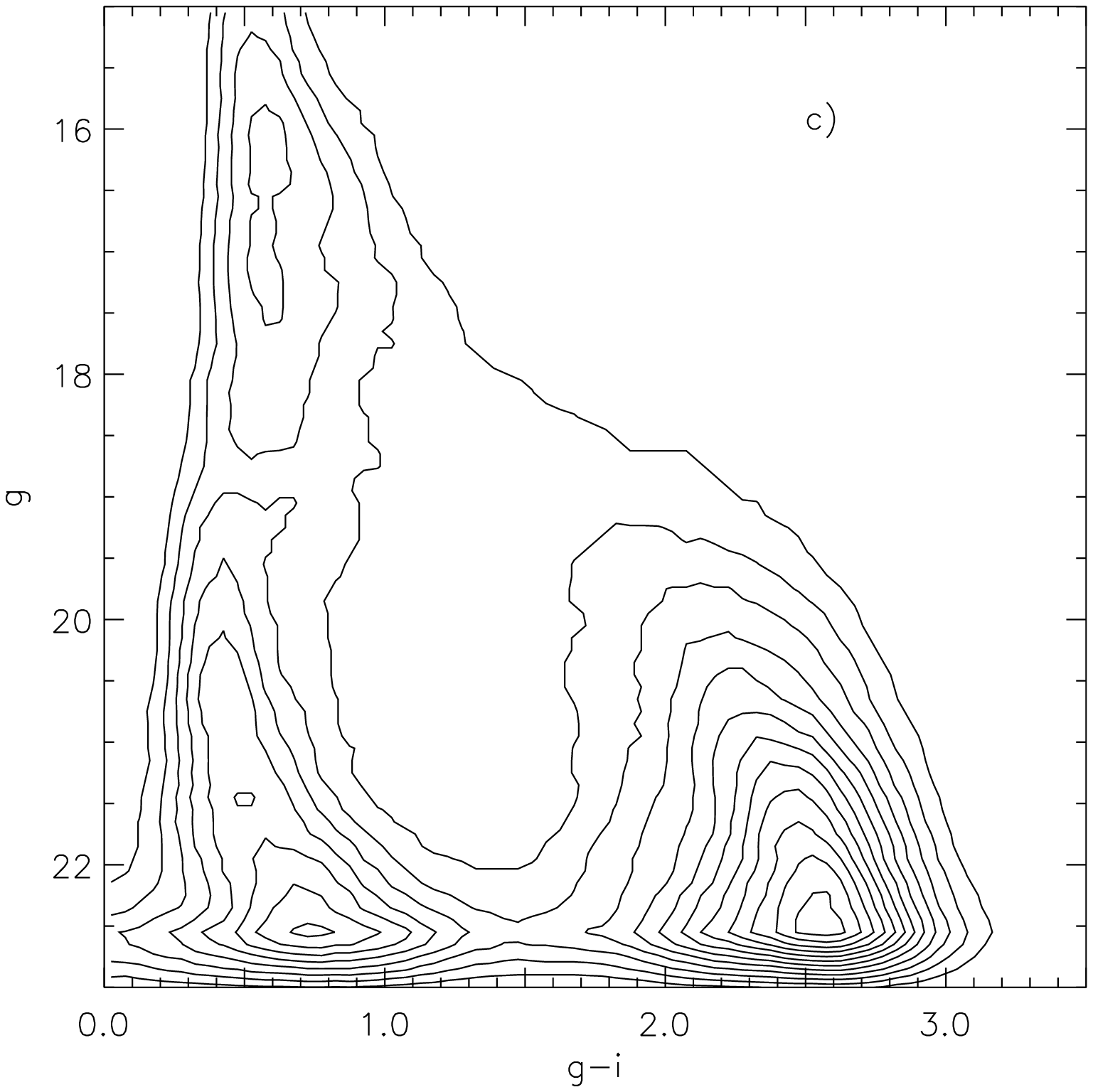}
\includegraphics[width=3in]{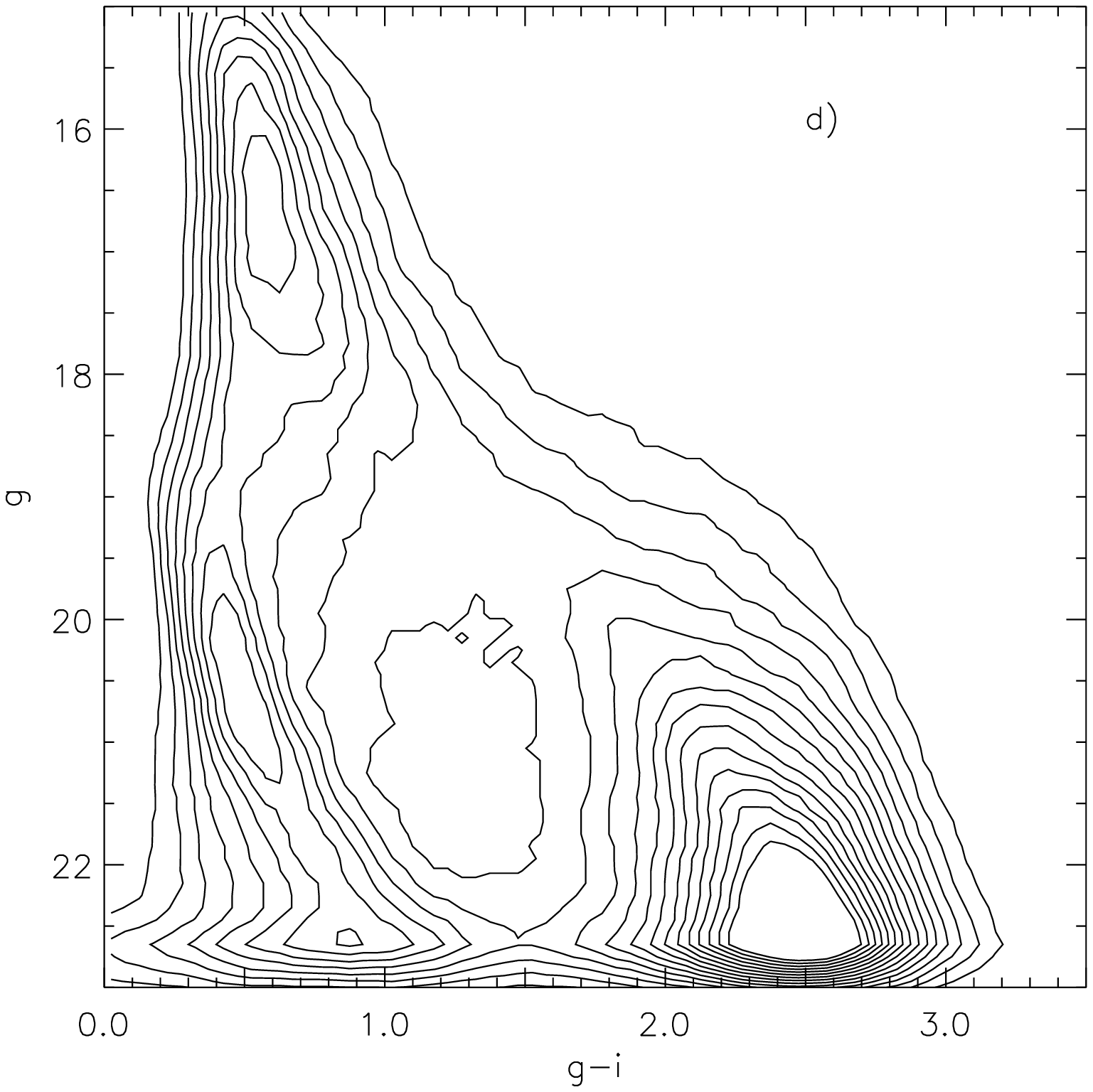}
\caption{The background filters $n_{bg}(g-i,g)$ of four $5^\circ\times5^\circ$ regions in our sample. The (RA,Dec) centers of each region are a: (138.5,17.5), b: (127.5,17.5), c: (138.5,5.5), d: (127.5,5.5). Region (d) greatly overlaps the missing data seen in Figure \ref{field}, but the qualitative features seen in panel (d) above are sufficient for our needs. Contours in all four panels are in steps of $\frac{1}{2}\sigma_a$ above the mean level of panel (a) to demonstrate the increasing stellar density at lower galactic latitudes. Specific features are discussed in the text.}
\label{nbg4}
\end{figure}

The same task must be carried out for the field to describe the background filter.  It is clear from Figure \ref{field} that the field does not have a uniform density of sources in our sample. Instead we see the density of stars rise at lower galactic latitudes, near the galactic plane. To investigate the effect the changes in the field star population would have on our analysis, we sampled the background population in several regions. Figure \ref{nbg4} shows the background $(g-i,g)$ CMD in four quadrants of the field. Other than the number of stars in each quadrant, the differences between these four CMD samples is minor, and we believe the changes in the CMD will not greatly affect our analysis, as was also concluded in R02. Thus we make the assumption that the background has a uniform CMD structure across the field studied, and that the stellar density may be described by a low-order surface fit, as described in \citet{2006ApJ...639L..17G}.

\begin{figure}[]
\centering
\includegraphics[width=3in]{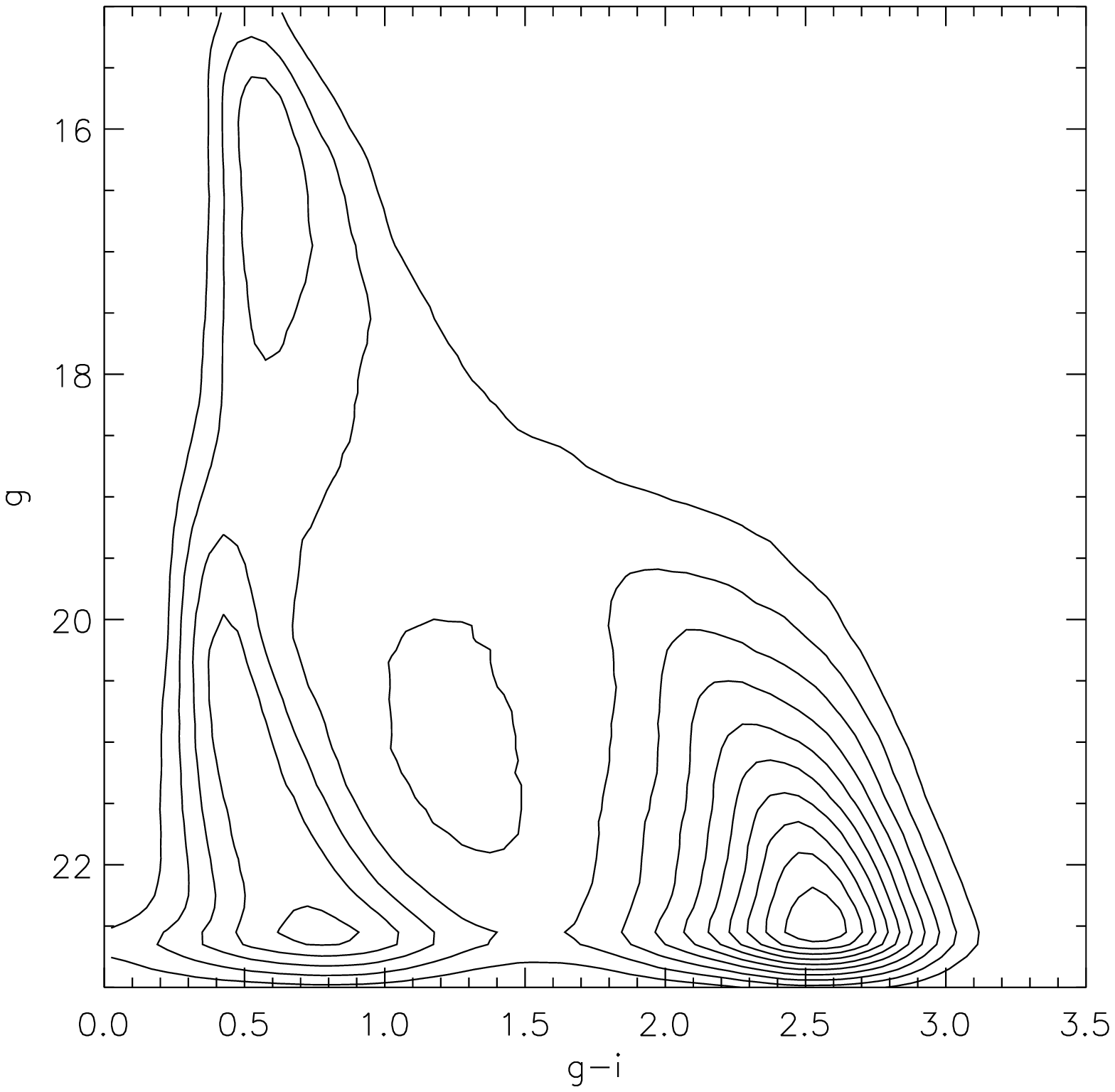}
\includegraphics[width=3in]{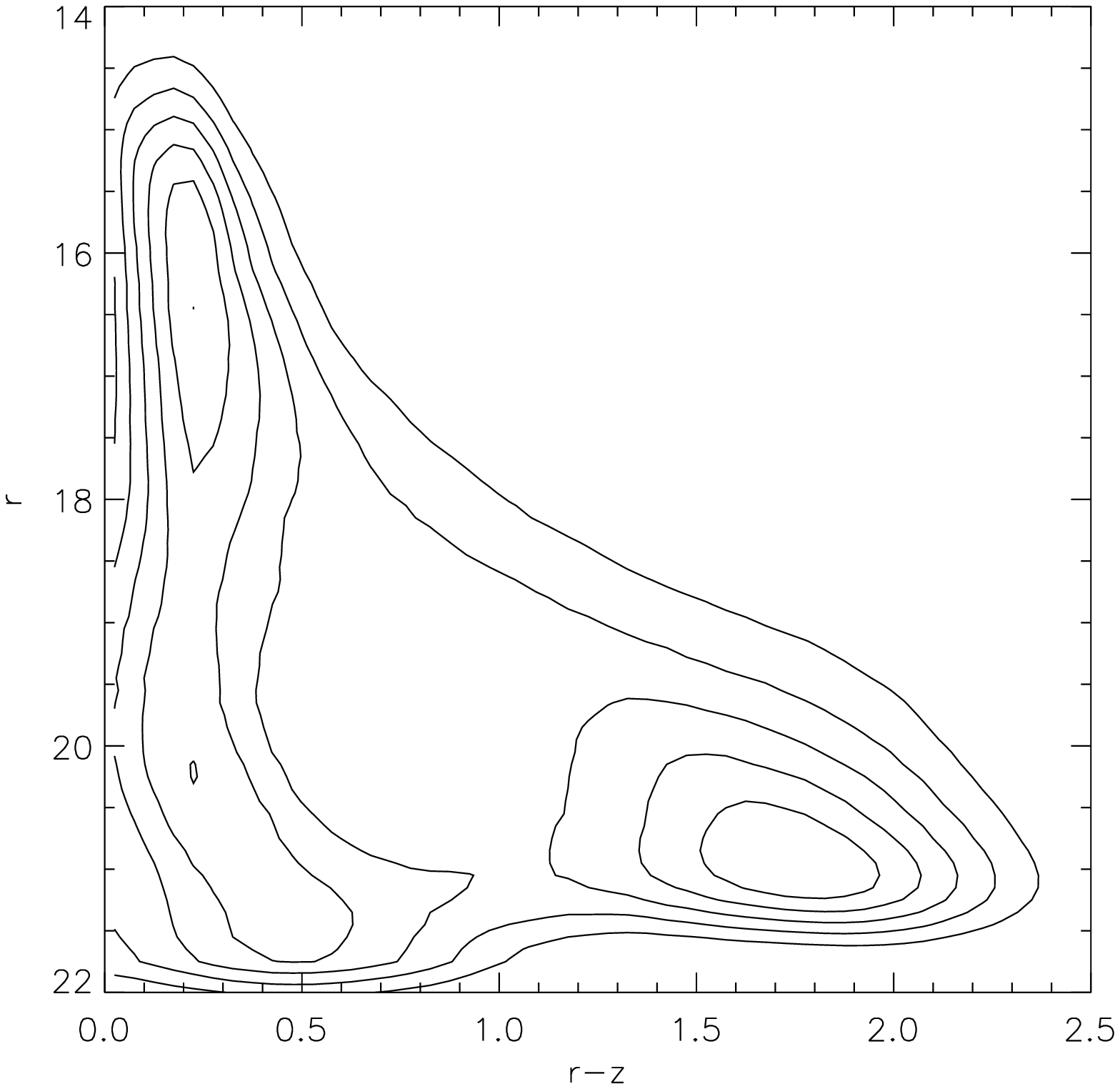}
\caption{The smoothed background CMD filters, $n_{bg}(g-i,g)$ (left), and $n_{bg}(r-z,r)$ (right). The sample is drawn from a region around M67 with 1$^\circ<$ radius $<8^\circ$. The contours are in increments of $\frac{1}{2}\sigma$ above the mean. Specific features are discussed in the text.}
\label{nbg}
\end{figure}

The background filter is referred to as $n_{bg}(color,mag)$, which, given any area on the sky provides the number of background stars in each color vs. magnitude bin within that solid angle. It is valid to use $n_{bg}$ and not $f_{bg}$ here because the background is a continuous and relatively smooth population, and the number of stars in any area can be estimated. By using all of the stars around M67 with a radius  between $1.5^\circ$ and $8^\circ$ (1,011,776) our background population is very well sampled. This annulus was chosen to avoid both the cluster and any potential tidal features, and the large hole in the data set in the south-east corner. Applying a boxcar smoothing algorithm and dividing by the area used, we create the two $n_{bg}$ filters shown in Figure \ref{nbg}.

Because the field population does not have a uniform distance or age, the familiar features of an ideal CMD (e.g. main sequence, turn off, red giant branch, etc) become severely blurred. The resulting color vs. apparent magnitude diagram is known as a Hess diagram \citep[e.g. see][]{macho}.
In the Hess diagrams in Figures \ref{nbg4} and \ref{nbg}, several distinct features are visible in the CMD contours.  The peak at $(g-i,g)=(0.6,17)$ is from the galactic disk (the thick disk according to R02). This feature is strongest in Figure \ref{nbg4}d, which has the lowest galactic latitude and thus should contain the most disk stars. The smooth feature at $g-i \sim 0.5$ is the main-sequence turnoff for the halo population, which is naturally most prominent in Figure \ref{nbg4}a, at the highest galactic latitude of our sample. The large red buildup of stars at $g-i \sim 2.5$ is the local field K and M-dwarf population, most of which are likely part of the disk. This is also most prominent in Figure \ref{nbg4}d.
 We have enforced magnitude cuts at both the bright and faint limits of our SDSS sample, and thus the $f_{cl}$ and $n_{bg}$ distributions have steep declines near the limiting magnitudes.

\begin{figure}[]
\centering
\includegraphics[width=3in]{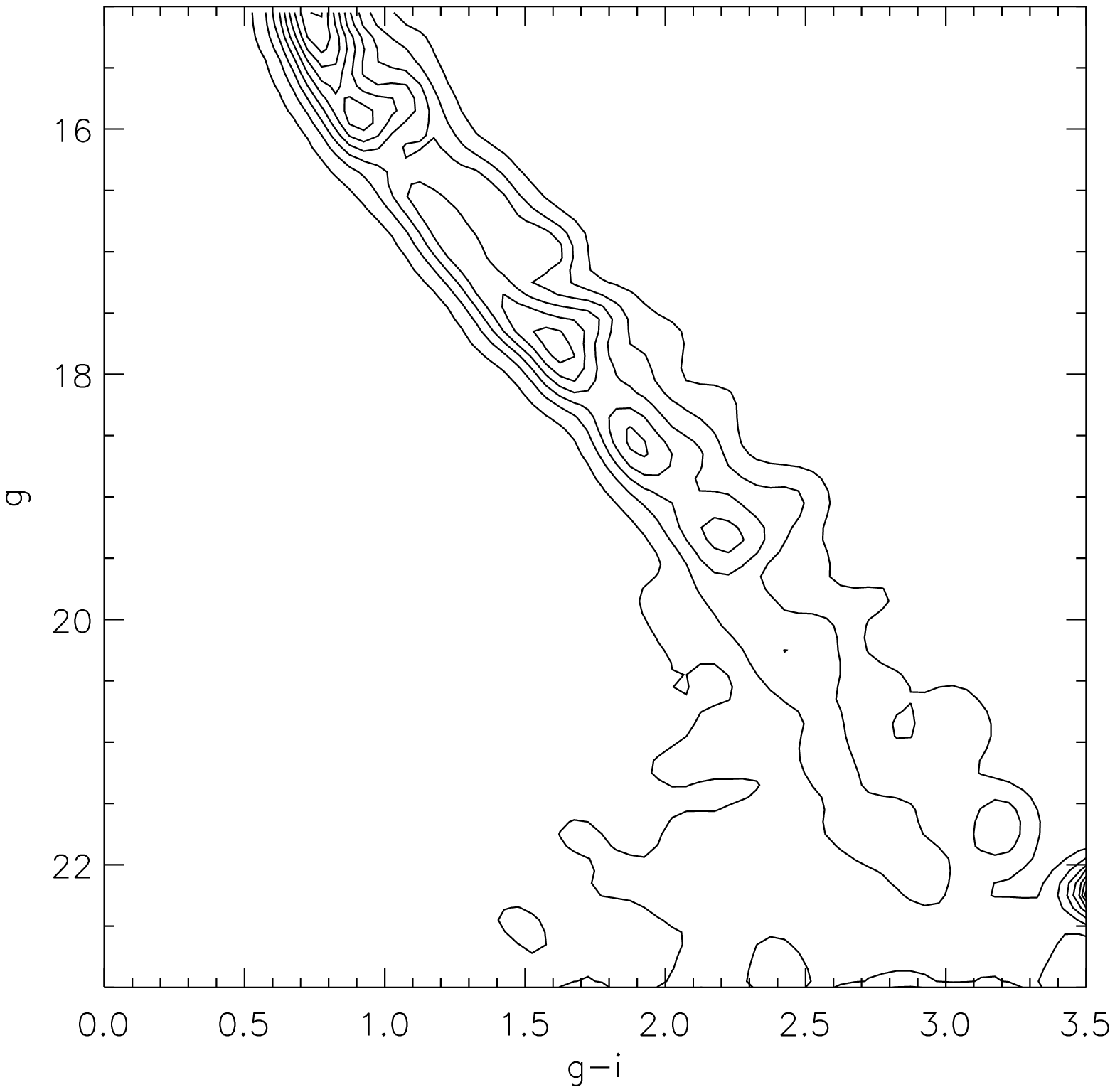}
\includegraphics[width=3in]{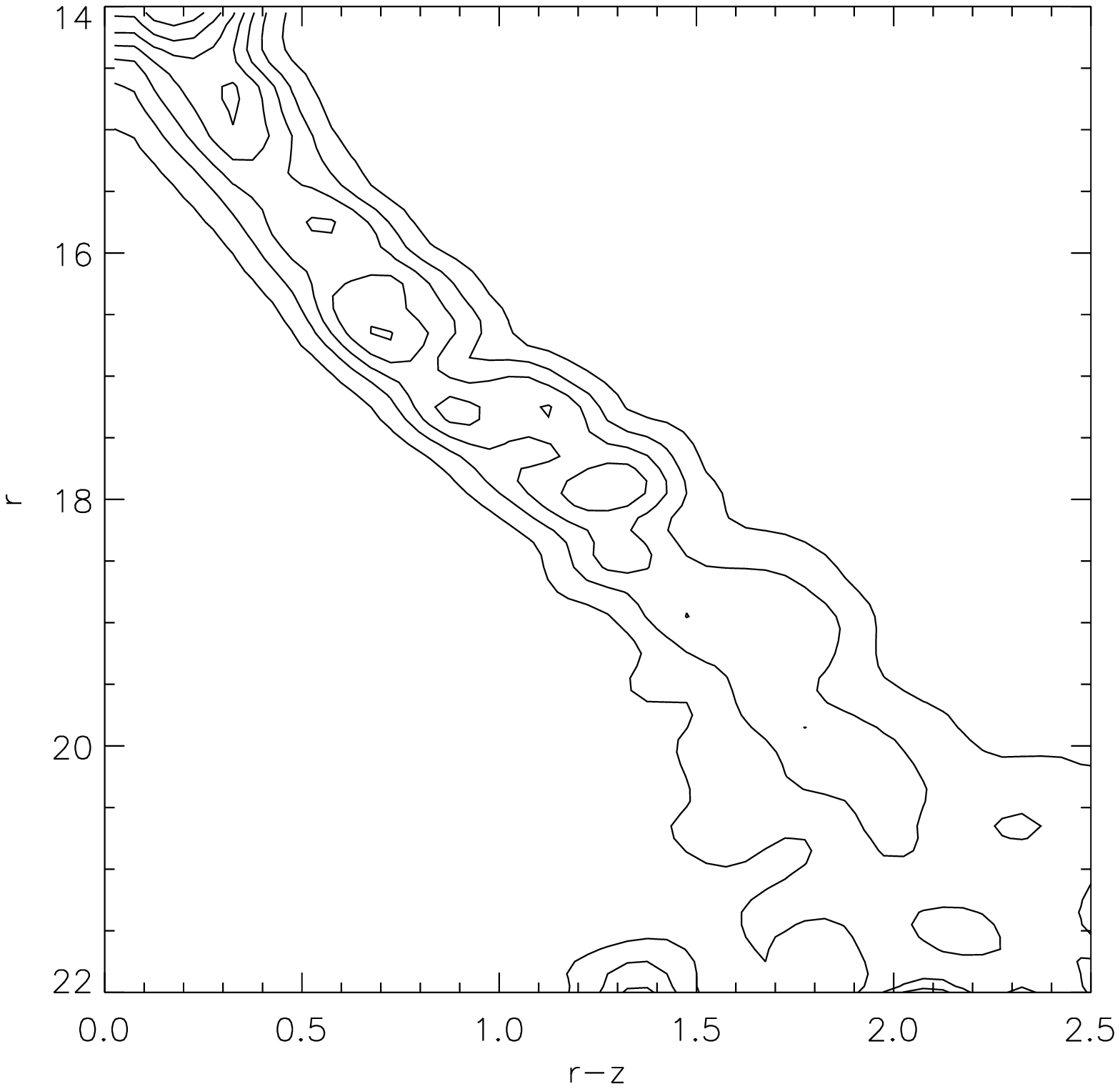}
\caption{The matched CMD filters, $h(g-i,g)$ (left), and $h(r-z,r)$ (right). The contours are $1\sigma$ increments above the mean of the $h$ functions.}
\label{h}
\end{figure}

To produce the optimal contrast, the normalized cluster filter is divided by the scaled background. This yields
$h=\frac{f_{cl}}{n_{bg}}$,
 which is shown in Figure \ref{h}. This produces a CMD matched filter that describes the stars which stand out from the background the strongest. For any given star, its color and magnitude will yield a value of $h$ related to its probability of membership in M67. Naturally this probability function will promote regions of the CMD where there are many M67 members but few field stars, and conversely punish stars with CMD positions having large field populations. 
 
 The spatial distribution of the cluster is then mapped by summing the values of $h$ inside small ($0.1^\circ,0.1^\circ$) spatial bins. We complete this entire process independently for both the $(g-i,g)$ and $(r-z,r)$ CMDs, and co-add the results, as per \citet{22deg}. While the probability distribution $h$ produces the maximum likelihood indicator for the stars, corrections for the field response to the $h$ filter must be done. To solve for the actual number of cluster stars found in each spatial bin, $\alpha$, the following formula from R02 is used:
 \begin{equation}
 \alpha = \left\{\sum\left[\frac{f_{cl}}{n_{bg}}\right] - \int f_{cl} d(color,mag)\right\} /\left\{ \int\frac{f_{cl}^2}{n_{bg}} d(color,mag)\right\}\,,
 \end{equation}
where $\frac{f_{cl}}{n_{bg}} = h$ is summed for all stars in a spatial bin as mentioned above, the integral of $f_{cl}$ accounts for the background response in a solid angle $d\Omega$, and the integral of $\frac{f_{cl}^2}{n_{bg}}$ is the signal-to-noise response of the filtering. This equation is the basis for our analysis with both the SDSS and 2MASS data sets, as well as the Monte Carlo simulations described in the following section.

\subsection{Analysis of Model Clusters}

In order to test the robustness of the matched filter method, we ran our algorithm using artificial clusters of known stellar composition and structure. By characterizing the efficiency and reliability of the algorithm we were able to examine issues of biases and errors for our results in \S4. To accurately measure the efficiency and determine the greatest sources of error, we inserted artificial clusters at random locations within our field. These tests were run $1000$ times on our SDSS data set in order to determine the reliability and significance of low-contrast features seen in \S4. 

The simulated clusters were created with a large range of core densities. The number of members was randomly chosen between $450\le N_{STARS} \le 1350$, while the core radius was randomly selected between $0\farcm9 \le R_{core}\le 22'$. This wide range was used to fully explore the densities and spatial compositions which open clusters might have.

The spatial composition of each cluster was created by randomly generating radial positions for every member, each being drawn from a gaussian profile with a standard deviation radius of $R_{core}$. The angular positions were then chosen randomly between $0^\circ$ and $360^\circ$, creating a roughly uniform circular distribution of stars. No tidal elongation was explicitly included, however the gaussian distribution of stars did at times produce asymmetric halos at large radii consisting of up to a few dozen stars.

The $griz$ magnitudes were formed using the handmade splines fit to M67 in \S3.1. A luminosity function for the $g$ band was used which had the form $g=(g_{*}^{0.77})\times8+15$, where $g_{*}$ is an array of $N_{STARS}$ elements filled with random values between 0 and 1. This function was chosen to provide an increasing number of stars with decreasing luminosity, but to have a deficiency as compared to the field star population of low-mass stars, as observed in old open clusters.  We also found that it approximately reproduces the observed luminosity function of M67. An equal mass binary sequence was created by increasing the magnitudes of a subset of stars by 0.75 mag in every band. The fraction of equal mass binaries was chosen at random between 10\% and 60\%.

Appropriate photometric scatter was created by sampling errors for the actual SDSS data. The mean error and the standard deviation of the errors were calculated for $griz$ bands in bins of 0.1 magnitudes. These errors were then mapped to the artificial stars by their corresponding $griz$ magnitudes, choosing for each star a random number within the gaussian error envelope. Adding these errors to the artificial photometry created a very realistic scatter about the M67 splines which increased with increasing photometric magnitude. 

For every model run, the simulated data was placed randomly in the field around M67, but was required to be greater than $1.5^\circ$ radially away from M67 to avoid cross-contamination, and less than $8^\circ$ away to stay well within the bounds of the SDSS data sample. The full matched filter analysis code was then run, sampling $f_{cl}$ from a $0.25^\circ$ radius around the model cluster and smoothed as before with the M67 $f_{cl}$. The background $n_{bg}$ was determined from an annulus around this with a radius between $1.5^\circ$ and $6^\circ$, encompassing $\sim$500,000 stars on average, but taking care to avoid M67.

\begin{figure}[]
\centering
\includegraphics[width=2in,angle=90]{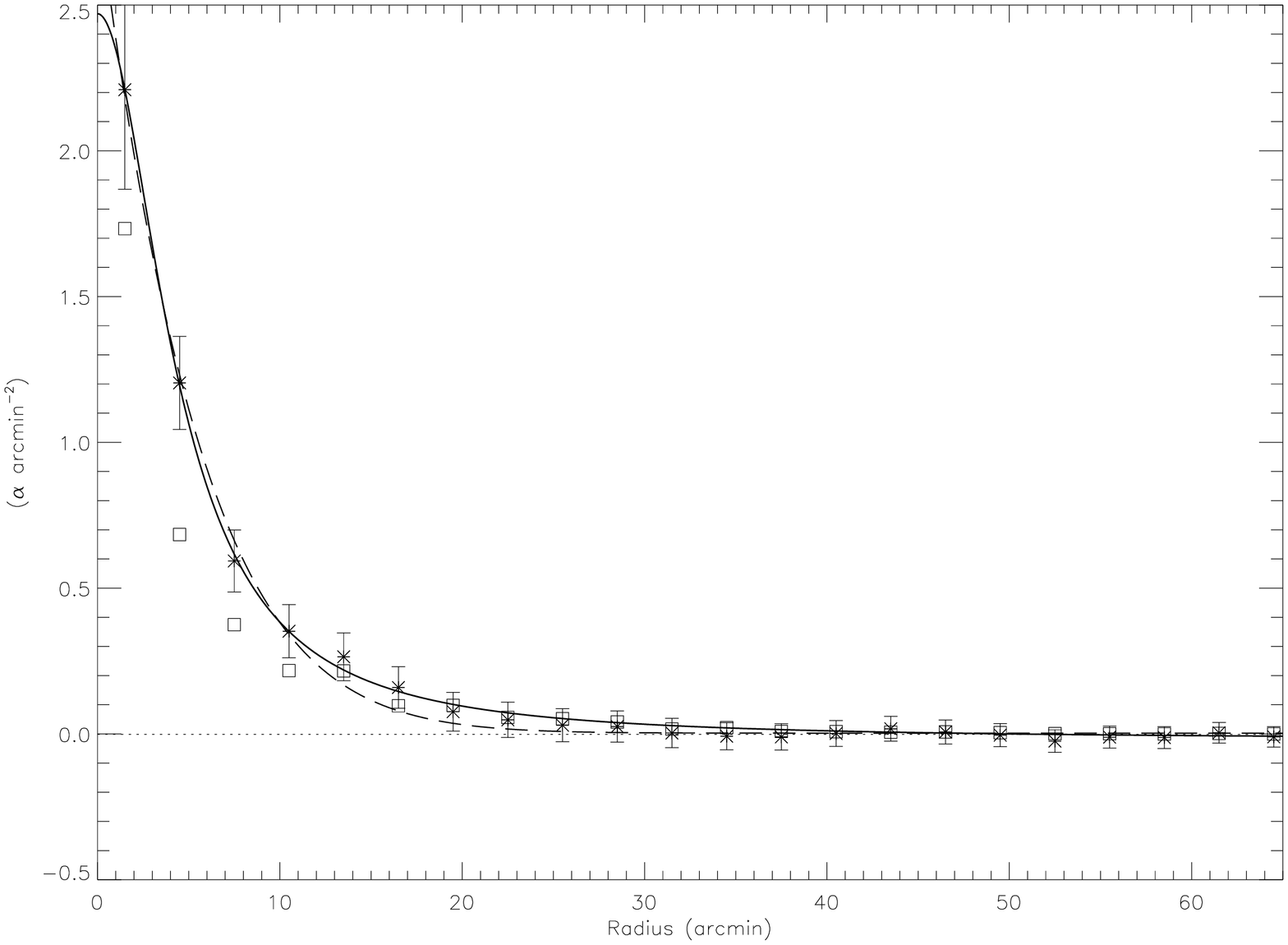}
\includegraphics[width=2in,angle=90]{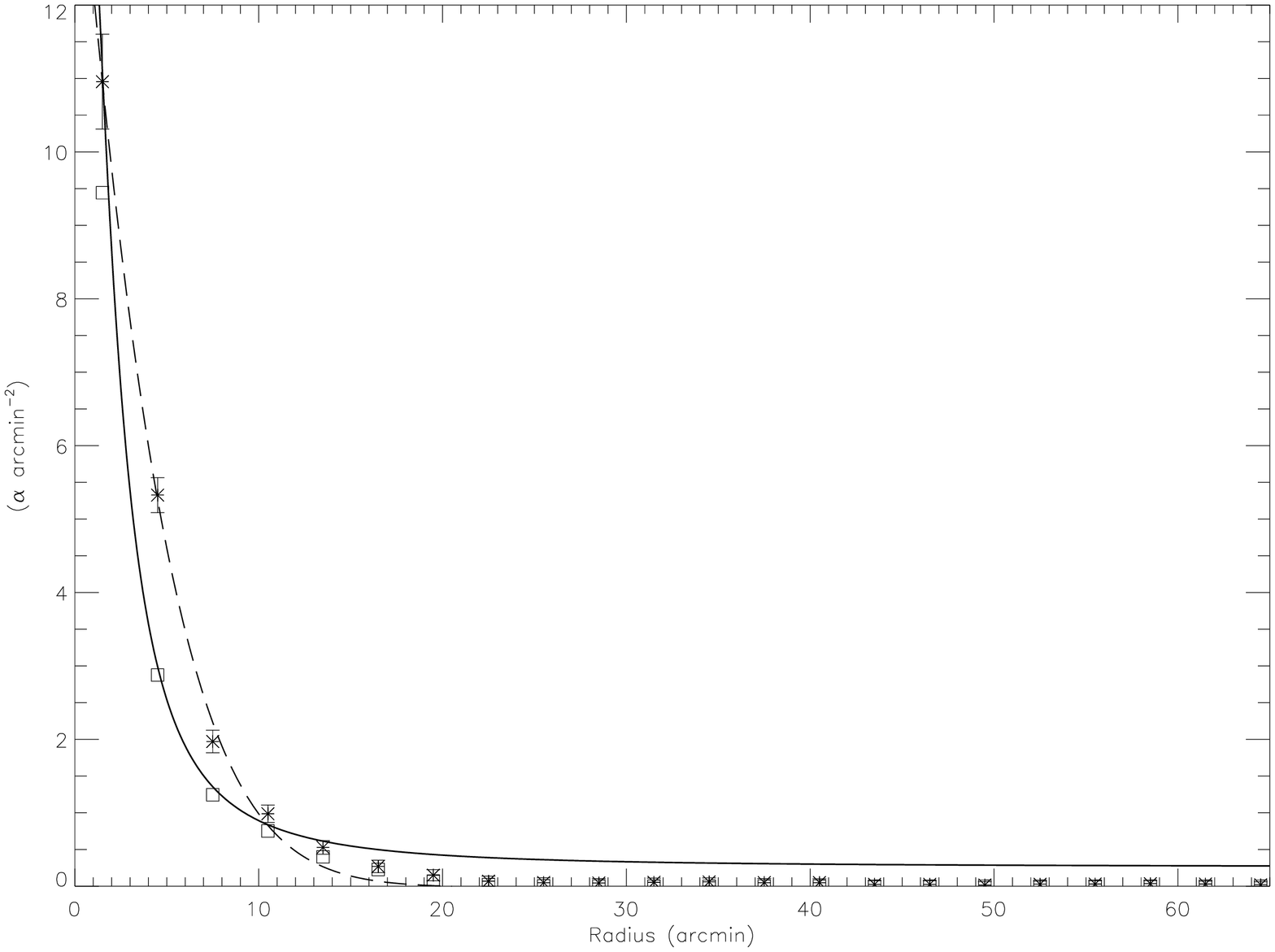}\\
\includegraphics[width=2in,angle=90]{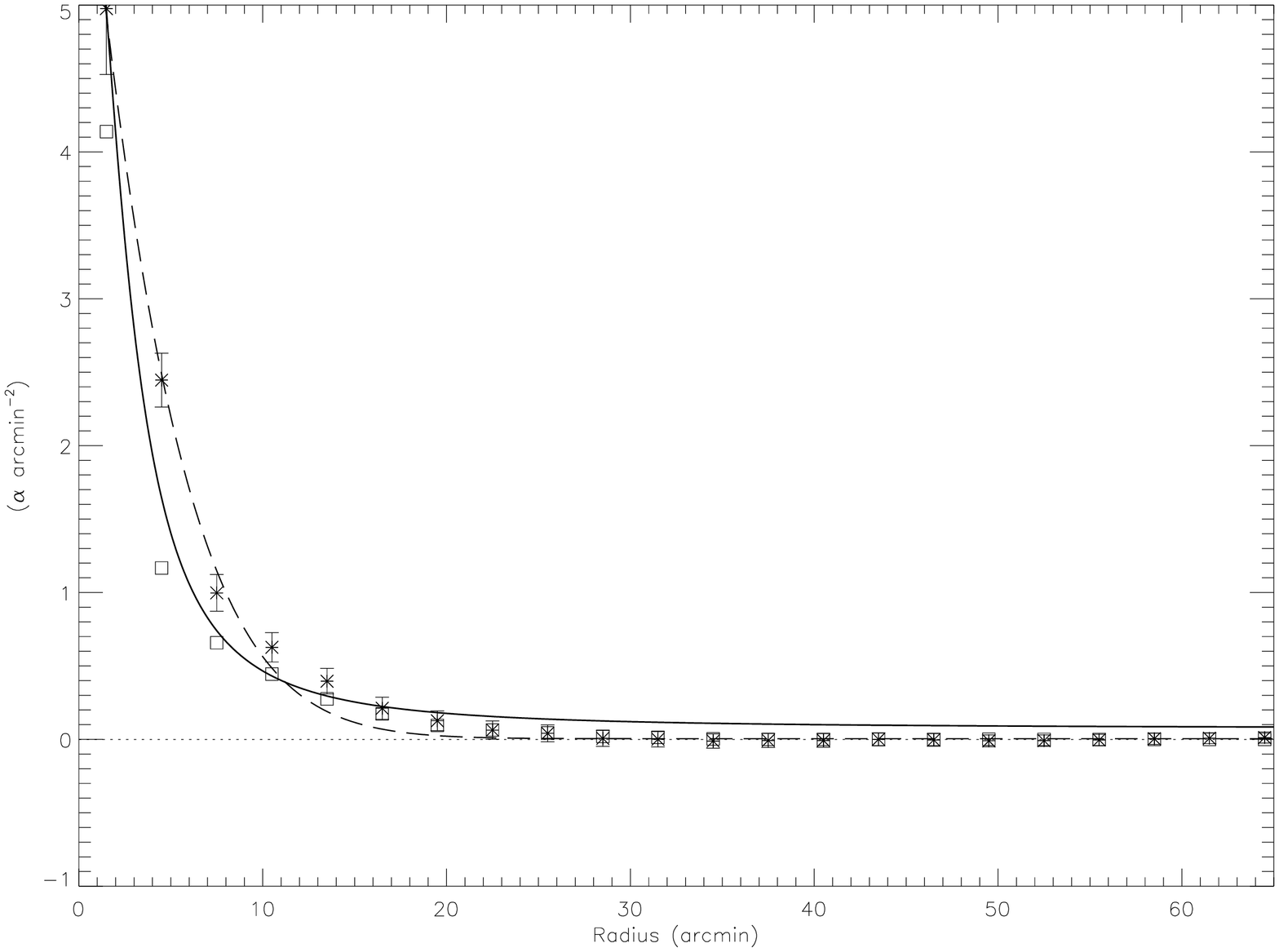}
\includegraphics[width=2in,angle=90]{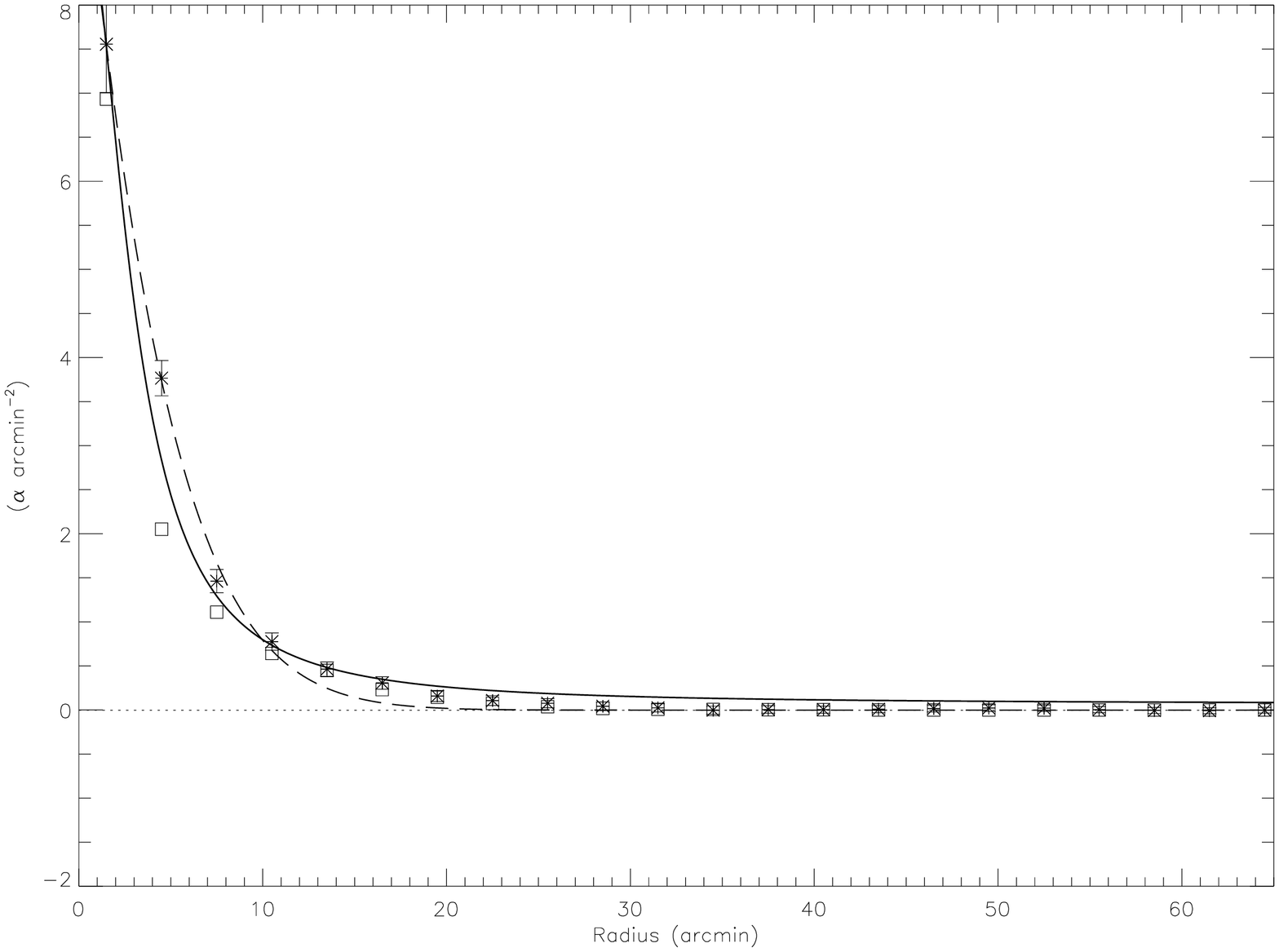}
\caption{The radial surface density profile for a random subset (4 of 1000) of our artificial cluster model tests. The solid line is a gaussian fit to the data, while the dashed line is a King-like profile fit. The dark stars are the recovered density at each bin, and the squares are the input model for this run. There is generally very good agreement  with the input data to the shape and distribution of the measured results, although we tend to overestimate the number of stars in the central bins.}
\label{model1}
\end{figure}

\begin{figure}[]
\centering
\includegraphics[width=2.5in]{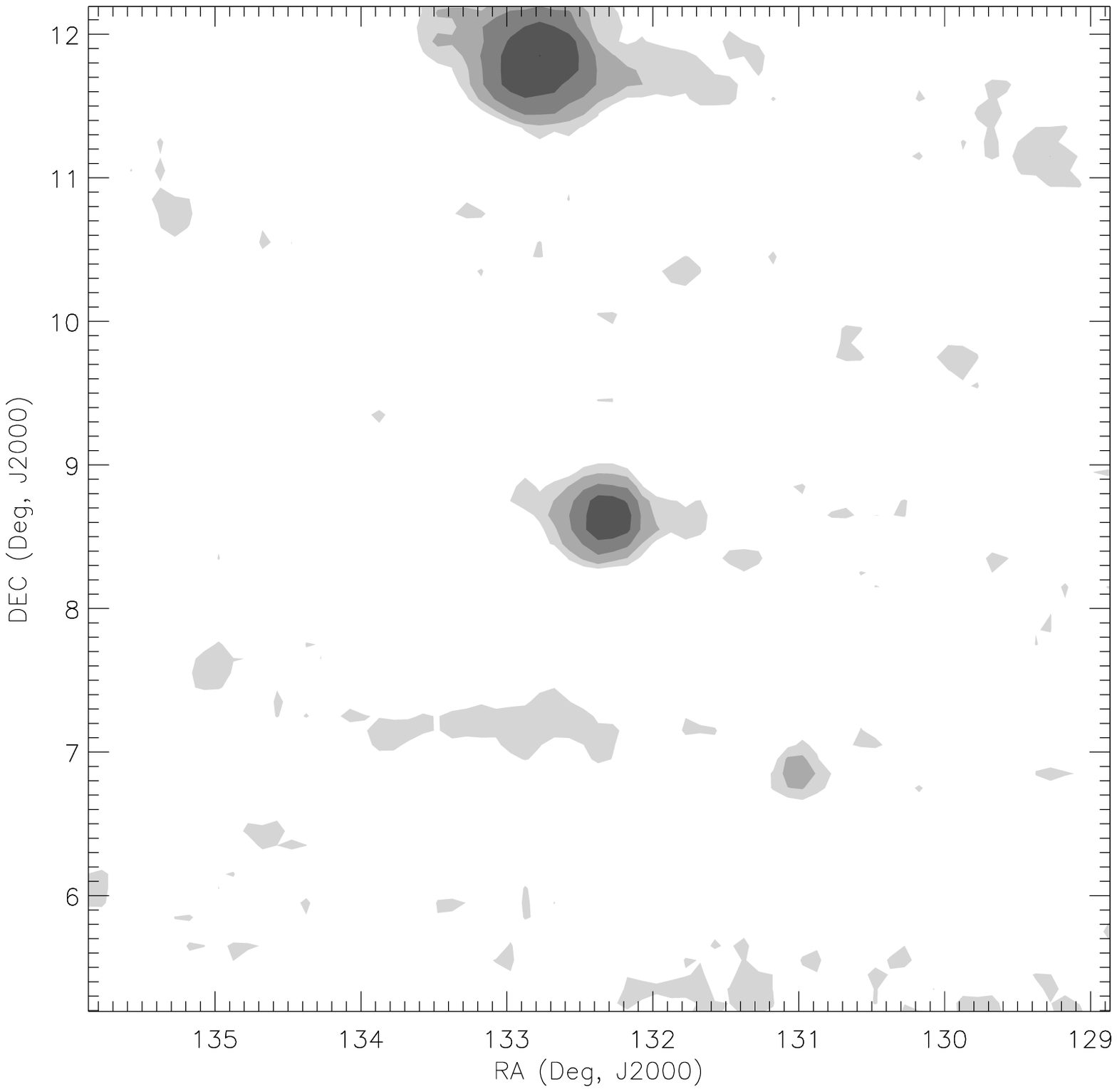}
\includegraphics[width=2.5in]{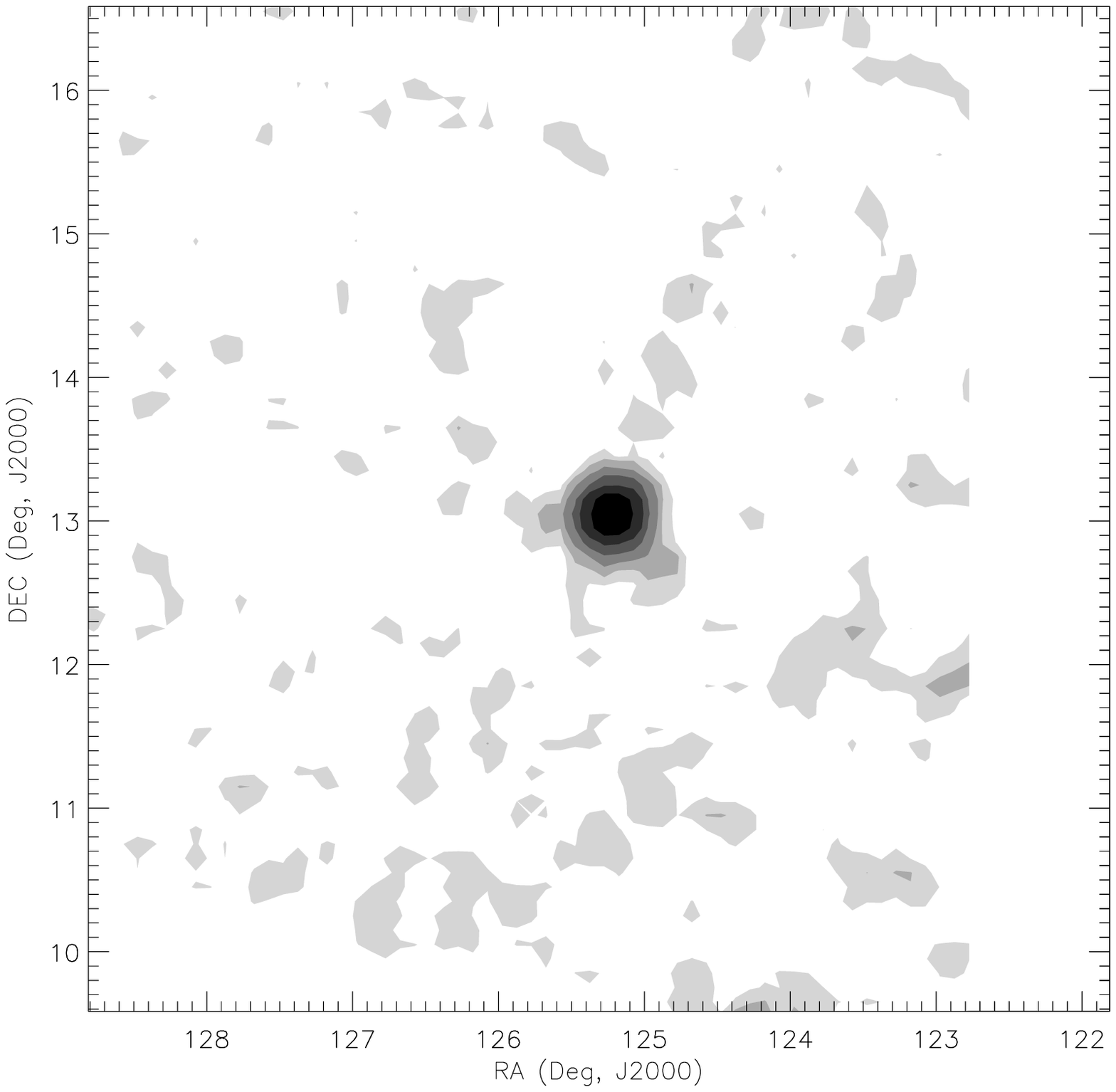}\\
\includegraphics[width=2.5in]{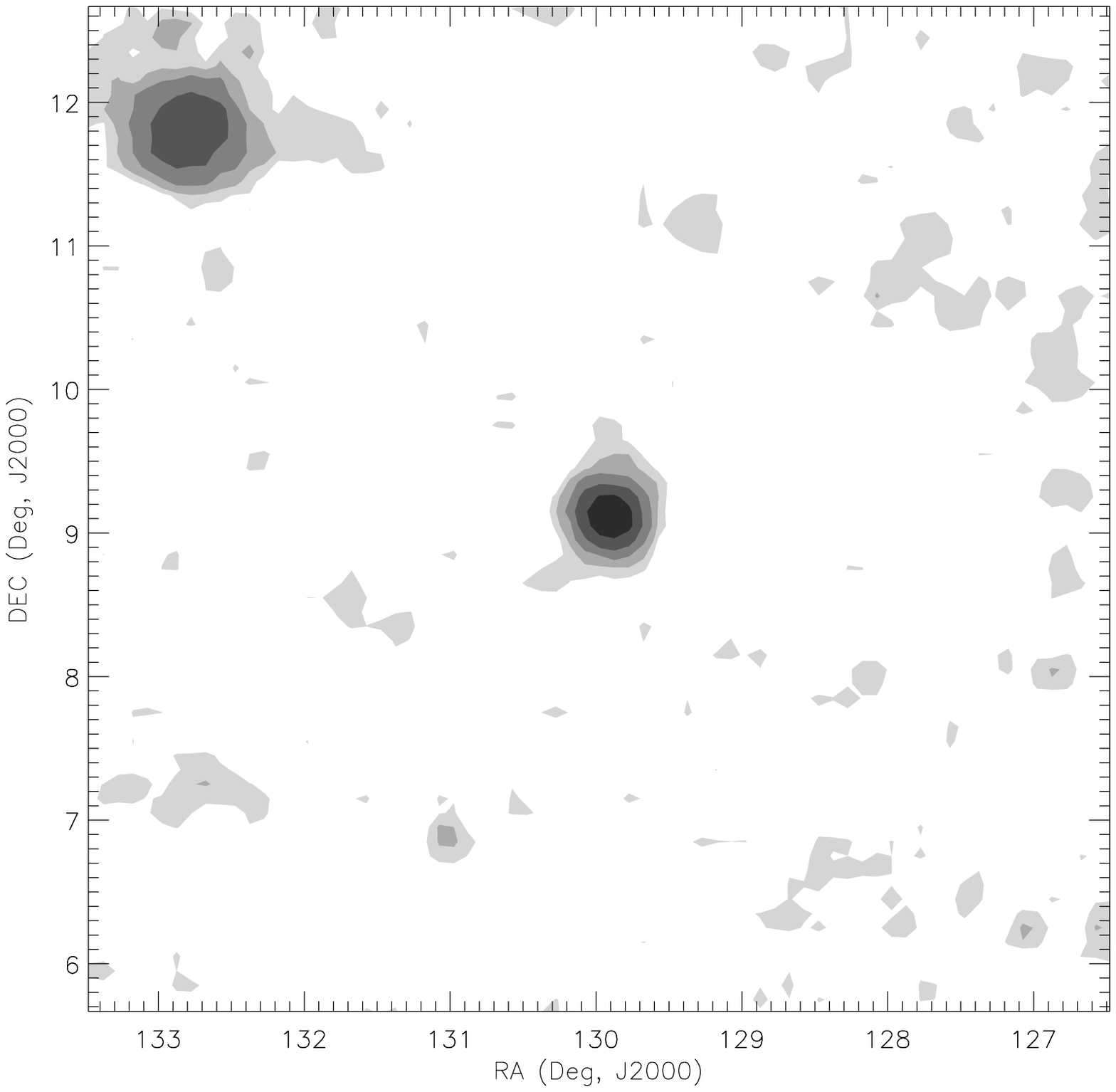}
\includegraphics[width=2.5in]{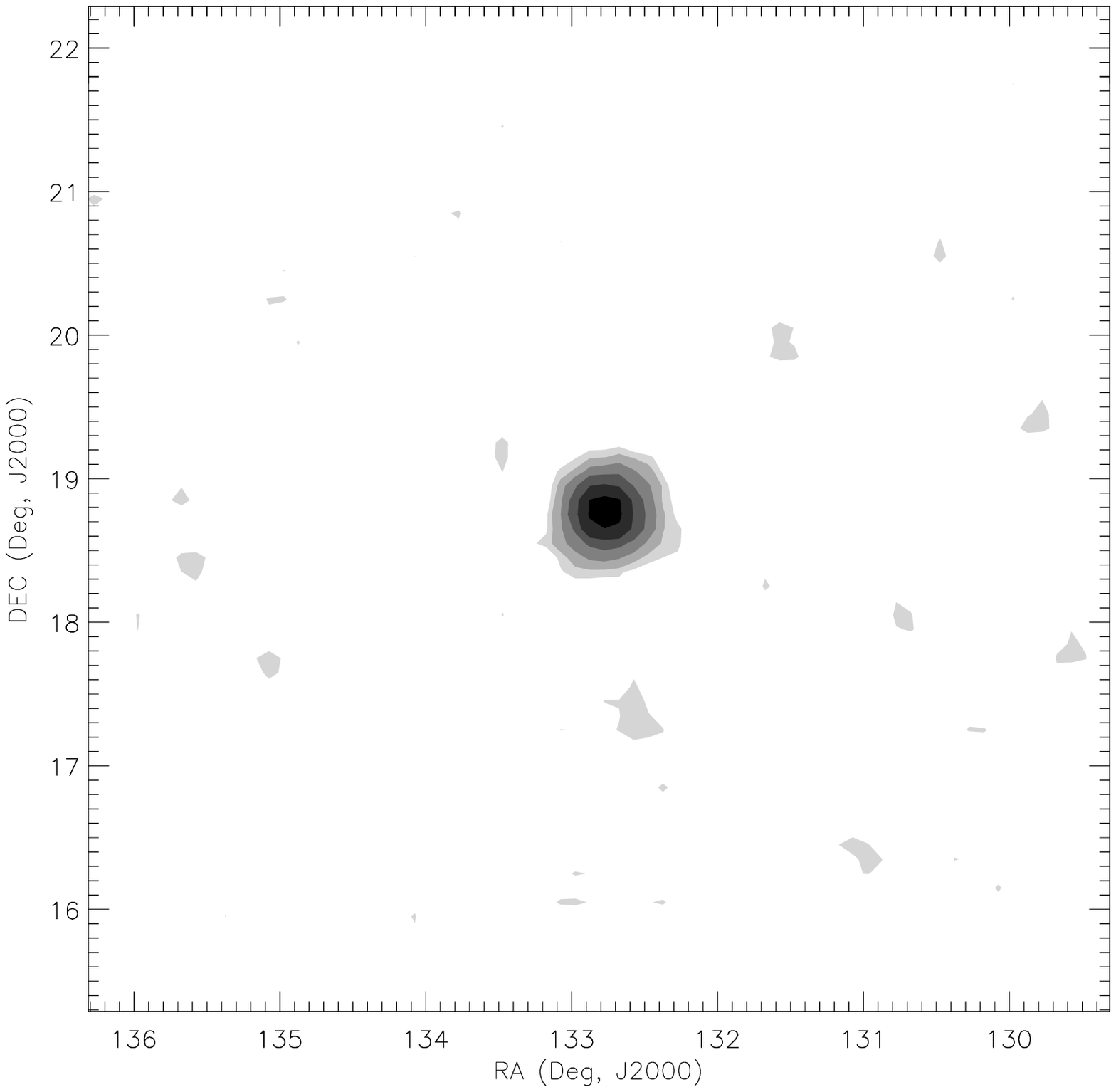}
\caption{The spatial map of our model cluster for a random subset (4 of 1000). These correspond to the radial surface profiles in Figure \ref{model1}. Contours are in increasing $\alpha$ (cluster stars), as defined for Figure \ref{map}.}
\label{model3}
\end{figure}

For every run of the simulation we recorded the simulated cluster position, surrounding background density, the estimated number of cluster members recovered by our algorithm, the radial surface density profile for each run, as well as every plot used in the analysis of M67. In Figures \ref{model1} and \ref{model3} we show example results for a random subset of our models. The resulting statistical analysis of the model runs show that we are able to reliably recover both the number of members and their locations. 

\begin{figure}[]
\centering
\includegraphics[width=5in]{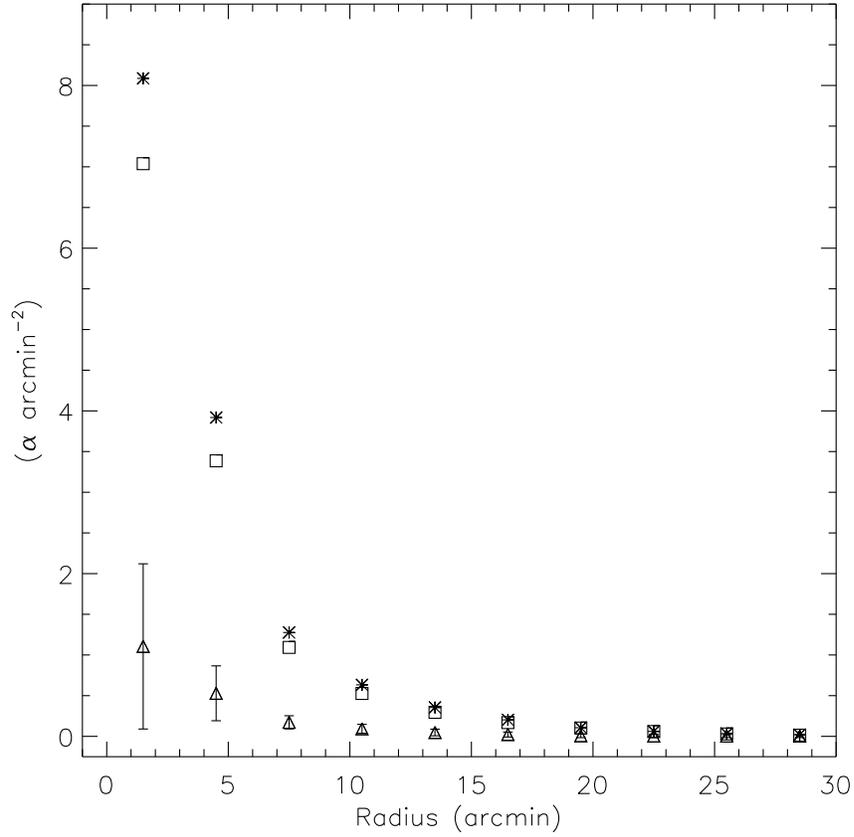}
\caption{The average radial surface density profile recovered from all 1000 of our models, shown as stars. The boxes are the average surface profile for the 1000 input models. Triangles represent the average residual between the input model and the recovered profile. Error bars are the standard deviations in each residual bin. }
\label{modelmed}
\end{figure}

Figure \ref{modelmed} shows the mean recovered radial surface density profile for our 1000 model runs, marked by stars. The squares show the mean radial surface density profile of the 1000 input models themselves. We are on average over estimating the surface density in the central bins, but are recovering the core radius well. Triangles mark the mean residual between the input models and the recovered profile. The error bars are one standard deviation in each residual bin. The small overestimation of the radial profile in the nucleus is attributable to the spatial smoothing kernel which will indicate more stars in the central radial bin than are actually observed. 

In every run of our models the radial surface density profile is fit with a least-squares gaussian profile. From the $1000$ model profiles we found on average our algorithm underestimated the core radius by 11\%, and after correcting this bias we found a standard deviation of 0\farcm60 for $R_{c}$. The number of stars detected by our algorithm was larger than the input models by, on average, 17\%, and we computed a standard deviation on the number of stars recovered to be 67 stars. We also determined that the input binary fraction was 1.27 times more than our recovered binary fraction, and found a standard deviation of 6\% for the resulting binary fraction.
None of the recorded statistics changed as functions of the background density, suggesting that our matched filter algorithm consistently was able to remove the field contamination.  We are attempting to improve these biases for future implementation of our algorithm.

Since each model run was required to be at least 1.5$^\circ$ away from M67, in some cases the artificial cluster can be close enough to M67 for it to be present on the radial profile plot at it's furthest extent. This can be seen in Figure \ref{model3}a for instance. These examples provide a useful benchmark for the robustness of the matched filter. Despite low-order changes in the local background population surrounding each run, and the variations in the model $f_{cl}$ compositions and model core densities, M67 is always nearly perfectly recovered. The elongated features of M67 seen in these examples match those found in the analysis of the following sections.

Visual inspection of a subset of the models, as demonstrated in Figure \ref{model3}, showed no examples of spurious extra-tidal features which deviated from the gaussian distribution caused by fluctuations in the background population. Any elongated features appeared to be caused by the input model. Further, our algorithm itself, while slightly overestimating the number of stars in the clusers' core, does not produce random cluster halo signatures. We therefore conclude that the matched filter method is effective in characterizing and removing the field star population and providing an accurate representation of the cluster within the range of our sample. Any extra-tidal structures or elongations found in the vicinity of M67 are therefore considered to be real and intrinsic to the cluster.

\section{Results}
\subsection{M67 Properties}
By analyzing SDSS data for M67 using the technique outlined in \S3.1, we were able to map the cluster distribution to a larger radial extent than has been done before. Figure \ref{map} shows our spatial map of the open cluster M67. A small ($0.1^\circ$) boxcar smoothing kernel has been applied to the summed $\alpha$ data, and a low-order surface fit to the background was subtracted to remove any residual large-scale variation of the field density. The core of M67 is strongly detected, and shows a circular distribution. The over-plotted circle, $25'$ in radius, denotes the furthest radius which BB03 detected the cluster against the surrounding field population using 2MASS data. There is however a significant low density asymmetric halo of stars well outside the core. The contours from light to dark are increasing levels of density. They are defined by the equation $level_j  = MED(\alpha) + (\sigma_\alpha\times2^j)$ where $MED(\alpha)$ is the median value of $\alpha$ over all spatial bins, $\sigma_\alpha$ is the standard deviation of $\alpha$, and $j\equiv\{1,2,3,4,5,6\}$. To ensure that $\alpha$ was well characterized, we created a histogram of $\alpha$ from every $0.1^\circ\times0.1^\circ$ spatial bin in our SDSS sample. This is shown in Figure \ref{alphabkgd}, and appears gaussian in shape and is centered around $\alpha =0$. Since our contours begin at two standard deviations above the peak shown in Figure \ref{alphabkgd}, and noting the reliability determined in \S3.2, we argue that these features, discussed in \S4.2, represent real structure around M67 which has previously gone unseen.

The radial surface density profile of M67 is shown in Figure \ref{surface}. The strength of detection in each radial bin is the sum of $\alpha$ values in that bin, divided by the area that the bin encompasses. This is directly analogous to the spatial map in Figure \ref{map}, and the units of $(\alpha$ $arcmin^{-2})$ are equivalent to $(stars$ $arcmin^{-2})$. The extended halo seen in Figure \ref{map} is visible from a radius of $\sim25'$ to $\sim50'$ where it falls to background levels. 

We fit this surface density profile using a King-like profile \citep{king62} to our data, which has the form
$
n(r) = n_{bkgd} + \frac{n_0}{1+(r/R_{c})^2},
$
where $R_c$ is the core radius, $n_{bkgd}$ the background surface density (not to be confused with the background filter $n_{bg}$), and $n_0$ the central peak surface density. This surface density profile equation was used perviously by BB03 to model M67.
This fit is shown in Figure \ref{surface} as the solid lines. Our King model fit provides $R_c=8\farcm24$ after correcting for biases from \S3.2, somewhat larger than the previous 2MASS based measurement of $R_c=4\farcm86$ by BB03. From our testing with the model cluster, we estimate the error to be $0\farcm60$. We believe the difference between our radial profile and that of BB03 is due primarily to the lower limiting mass our study probes. We do give a direct comparison to BB03 by using the matched filter algorithm on the M67 2MASS data and provide more discussion in \S4.3. After fitting the first-order King model used in BB03, we attempted to determine the tidal radius for M67, $R_t$.  Using the equation
$
n(r) =  n_{bkgd} +n_0\{[1+(r/R_c)^2]^{-1/2} - [1+(R_t/R_c)^2]^{-1/2}\}^2
$ from \cite{king62} and the values we determined for the first-order fit above, we found the estimated tidal radius for M67 to be $R_t = 64'\pm21'$. Errors here are determined from standard deviation of the RMS scatter on our fit. This corresponds to a radius of 16.8 pc at the distance of M67, much larger than the determination using bright stars in M67 by \cite{2007A&A...468..151P}.
By summing the number of stars in each radial bin in Figure \ref{surface} we estimate the total number of visible M67 members to be $1385\pm67$, where our error is again adopted from our models in \S3.2.

\subsection{Composition of the Halo}
Having found evidence for a large halo around M67, we investigated the differences between the halo and core populations. Variations in the types of stars in the inner and outer regions of the clusters might suggest an origin for the elongated halo. In dynamically un-evolved systems, the initial stellar luminosity function rises towards lower masses.
BB03 use this property to investigate the evidence for mass segregation in M67. They created a set of 2MASS J-band luminosity functions in three regions (core, inner-halo, and outer halo).  A higher fraction of low-mass and faint J-band stars were seen in the outer halo, and thus mass-segregation is implicated.

Binary star systems are composed of two stars which are often too close to be visually resolved. This single point source is observed to be brighter than a single-star of the same apparent color. These systems also have higher mass than single-star systems by definition. The effects of mass segregation discussed above are therefore applicable to binary stars which act as more massive single stars. Previous cluster studies have observed a stronger concentration of binary stars in the cores than in the outer envelopes of older clusters  \citep[e.g.][]{fan96}.

The binary fraction and luminosity function for each region of our data can be probed in much the same way as the spatial structure. Rather than summing $\alpha$, the number of potential member stars, in each spatial bin, we summed $\alpha$ in three concentric regions for every magnitude bin (for the luminosity function) or across the main sequence (for the binary fraction).  These three regions were the core, halo, and a large background annulus centered around M67.

Calculating the binary fraction across the entire magnitude range requires properly collapsing the CMD into the color plane. We employed a technique used by \citet{clark04} to collapse the main sequence of the globular cluster Palomar 13. Since a spline has already been fit by hand along the M67 main sequence in \S3, we subtracted the main sequence spline color from every star's color, thus centering the main sequence on $\Delta Color = 0$.
Because the equal-mass binary sequence is 0.75 mag brighter than the main sequence, it's $\Delta Color_{bin}$ position was easily tracked. We then divided $\Delta Color$ by the $\Delta Color_{bin}$ of the binary sequence, creating a reduced color $R \equiv \Delta Color / \Delta Color_{bin}$. This placed the main sequence at $R=0$ and the equal mass binary main sequence at $R=1$. To calculate the binary fraction we summed all of the $\alpha$ values in each reduced color bin, with a range of $-1\le R \le 2$. The luminosity function is found by summing all of the $\alpha$ values in every magnitude bin, again cutting out stars outside the reduced color range $-1\le R \le 2$.

In Figs. \ref{core}, \ref{halo}, and \ref{bkgd} we investigate the M67 core, halo, and background populations respectively. The central panel shows the reduced CMD, where the darker pxels show increasing $\alpha$ sums in each CMD bin. The right panel shows the sum of $\alpha$ values along the magnitude axis, while the bottom shows the same along the reduced color axis. Since the equal-mass binary main-sequence was created by translating the main-sequence spline 0.75 magnitudes brighter, a discontinuity in the calculation of $\Delta Color_{bin}$ arises at the end points. Thus we have reduced our magnitude range by 0.75 on both the bright and faint end to avoid these biases.

The core population in Figure \ref{core} (radius $< 0.8^\circ$) shows a clear double gaussian peak along the reduced color $R$ axis, and a deficiency of faint (low-mass) members along the magnitude axis. The halo in Figure \ref{halo} (0.8 $<$ radius $<1.2^\circ$) contains a more flattened luminosity function, and a diminished secondary peak in the binary fraction. These two figures do not however properly account for the contribution the field population would have, especially in the halo which samples a larger area and has a lower expected density of cluster members. Thus in Figure \ref{bkgd} we made the same measurements for a much larger sample of stars ($1.5 <$ radius $< 6^\circ$) to estimate the field contamination in our binary fraction and luminosity function. This shows the response to the $h$ filter that the background has. 

The background contribution was normalized and then scaled by the area in the core and halo regions respectively. This bias contribution was then subtracted from the M67 core and halo samples. Figure \ref{bf} shows the background subtracted binary fraction estimations for the core and halo of M67. A gaussian profile was fit to the primary peak for $R<0.25$. The residual in the core was nicely fit by a second gaussian profile, centered at $R\approx1$. Integrating the two gaussian profiles, we measure a fraction of binary point sources to be $21$\% in the core of M67. This is comparable to the \citet{fan96} determination of 16\% and 22\% from \citet{1993AJ....106..181M}. Correcting for the artificial cluster tests in \S3.2 above, our estimation of the binary fraction is increased to 26\%. This is necessarily a lower limit on the fraction of binary sources as many stars may be binary systems without detectable flux excess. No significant gaussian binary star residual was found in the halo population, although the signal is noisier. Both the core and halo samples go to $\alpha =0$ for $R < 0$ and $R > 1$. The luminosity function is given the same treatment, and the background corrected functions are presented in Fig. \ref{lf}. The halo shows a more flattened low-mass contribution. The core however contains a significant lack of low-mass members, even after accounting for the background field contribution.

The dichotomy between the core and halo populations is consistent with qualitative ideas about mass segregation. As discussed in \S1, an old cluster such as M67 is expected to be losing low-mass members to the outer regions of the cluster, while the core becomes increasingly more concentrated with higher mass stellar systems. This simplification of course assumes a single static population of stars and also ignores the effects of classical stellar evolution, whereby higher mass stars die before lower mass, and skew the observed present day mass function. Assuming an age for M67 of $\sim$4 Gyr, it would be expected that late A or early F stars would be leaving the main sequence, and thus would be brighter than our limiting magnitudes with SDSS. We do not therefore expect stellar evolution to contribute significantly to the observed luminosity functions of our sample. \citet{2008ApJS..179..326A} investigated the effects of crowding and stellar density on the photometric completeness of the SDSS pipeline, including the M67 region. They found that the automated pipeline recovered a comparable number of photometric sources as compared to manual reductions of the imaging for the open clusters M67, NGC 2420, and NGC 6791. Thus we conclude that, while manual reductions of wide-field imaging such as in our sample would be preferable, the SDSS pipeline is able to reliably probe the regions around high galactic latitude open clusters.

\subsection{Comparison to 2MASS} 
To provide a complete comparison of our matched filtering method with previous results for M67, we ran our algorithm on $J, H,$ and $K_s$ band photometry from 2MASS. Using the 2MASS interface available on the internet entitled Gator\footnote{http://irsa.ipac.caltech.edu}, we retrieved a $10^\circ\times10^\circ$ box surrounding M67. This produced 264,354 point sources. The cluster CMD probability distributions $f_{cl}(J-H,J)$ and $f_{cl}(J-K_s,J)$ were drawn from all stars within a radius of $0.25^\circ$ around M67, and the background CMD from everything with a radius greater than $1.5^\circ$. Our IDL code for the SDSS data set was modified to use the 2MASS $JHK_s$ data, and the analysis was carried out to measure the spatial distribution of M67. We did not measure the binary fractions or luminosity functions for the halo and core populations with the 2MASS data.

M67 as seen by 2MASS contains a different range in spectral types than is observed by the SDSS, reaching significantly higher masses. This is shown Figure 2 of BB03, with the M67 CMD reaching a spectral type K0 at the faint limit, which we adopted to be $J=17$. In addition to the upper main sequence, the 2MASS CMD also contains the turn-off and red giant branch. The average stellar density in our 2MASS sample was 0.73 stars arcmin$^{-2}$, whereas the average in our SDSS sample was 1.4 stars arcmin$^{-2}$ over the same spatial area. 

Figures \ref{2mass1} and \ref{2mass2} summarize our 2MASS results for M67. These are analogous to Figures \ref{map} and \ref{surface} respectively. An asymmetric distributions of stars are seen in Figure \ref{2mass1} which extend beyond the detection of the BB03 study. The elongation seen in Figure \ref{map} is not well reproduced by the 2MASS data, however the mass ranges are considerably different.

Two significant differences between the SDSS and 2MASS results are apparent in Figure \ref{2mass2}: the 2MASS data yields a significantly higher surface density in the core, and the core radius fit with a King profile is much smaller ($R_c =4\farcm12$). This core radius corresponds very well with determinations by BB03 and \citet{fan96}.  
Because the 2MASS sample contains higher mass stars than the SDSS sample, the discrepancy with \S4.1 in core radius is expected. The higher mass stars in the 2MASS data will be more centrally concentrated in the core. These are too bright to be observed in the SDSS data and thus the SDSS will probe a mass range with a larger core radius.

Mass segregation differentiates all of the stars in a cluster according to their mass. The SDSS and 2MASS core radius comparison provides two estimates with a large separation of mass. To test that mass segregation was still evident in the 2MASS data alone we split our 2MASS sample into a bright bin and a faint bin. The separation was chosen to be $J=12.5$ which roughly corresponds to the main sequence turn-off. The full matched filter was rerun for each 2MASS subset. The bright and faint samples yielded core radii of $3\farcm70$ and $4\farcm67$ respectively, indicating that mass segregation among the high mass sample is clearly evident, as found in \S4.2 above.

The asymmetric halo around M67 shown in Fig. \ref{map} appears elongated roughly in-line with the proper-motion vector. \citet{fan96} have previously suggested an elongation of the core in M67 which roughly matches what is seen in Fig. \ref{2mass1}, but does not resemble what is seen in Fig. \ref{map}. 
A strong correlation cannot be made between the \citet{fan96} results and our work because \citet{fan96} considered asymmetries through 8 angular bins covering $45^\circ$ each. Their optical data was also limited to a photometric depth of $V=21$ and 0.3 magnitude errors making comparison to our SDSS data tenuous. Qualitatively the core radius and  tidal features of \citet{fan96} match our 2MASS analysis of M67. Our matched filtering has been able to reproduce known structure for M67 using 2MASS, and revealed a slightly increased radius of detection for the halo compared to BB03 due to our more complete subtraction of the contaminating field population.

\section{Discussion and Summary}

We have used a matched filter algorithm on SDSS photometry for identifying probable members of the open cluster M67. This study has revealed a core radius $R_{core} = 8\farcm24\pm0\farcm60$, and an asymmetric distribution of stars outside the previously known tidal extent. This asymmetric halo extends nearly $60'$ from the cluster center. The total number of stars measured within our SDSS data for M67 is $1385\pm67$.

We have also tested the reliability of our matched filter method using $1000$ artificial cluster simulations. In our models we tended to overestimate the number of members  by 17\% on average, and underestimate the core radius by 11\%. Overall the matched filter technique appears to be a robust method for identifying low-density stellar populations against a significant background. 

Our measurement of the binary fraction in the core of M67 is in good agreement with many previous studies \citep[e.g.][]{1993AJ....106..181M,fan96}. Correcting for systematic biases measured with our models, we estimate a lower-limit on the fraction of binary sources detected to be $26\pm6\%$. Since each binary source is composed of two stars, this corresponds to at least $45\%$ of all stars in our M67 core sample belonging to binary systems. This represents a lower limit as many of the ``single'' stars are possibly bound to very low mass stars which would place them blue-ward of the equal mass binary main sequence. We also do not correct for stars which may be obscured within the cluster. 
\citet{bb05} measured a binary fraction for the core of M67 to be $39\pm16\%$, which agrees with our determination within the respective errors. The increased central concentration of binary stars in the core, along with the greater fraction of low mass stars in the halo, fulfills the expectation by several studies that M67 has indeed undergone significant mass segregation. If N-body simulations for M67 \citep[e.g.][]{m67nbody} are to be believed, this cluster is near the end of its life as a gravitationally bound and distinct stellar population.

BB03 have recently carried out a very similar examination on M67 using 2MASS. The relatively bright limiting magnitude in the infrared meant that BB03 sampled higher mass stars than the SDSS. Our SDSS DR6 sample for M67 reaches a spectral type of M4 \citep{boo07}, but is limited at the bright end to mid K type stars. It is expected then that our SDSS sample would show signs of mass segregation to further radii, as we probe significantly lower masses. Since the elongated halo is not reproduced with our analysis of the 2MASS data, we suggest it must be primarily made up of the lowest mass members in M67.

\begin{figure}[]
\centering
\includegraphics[width=5in]{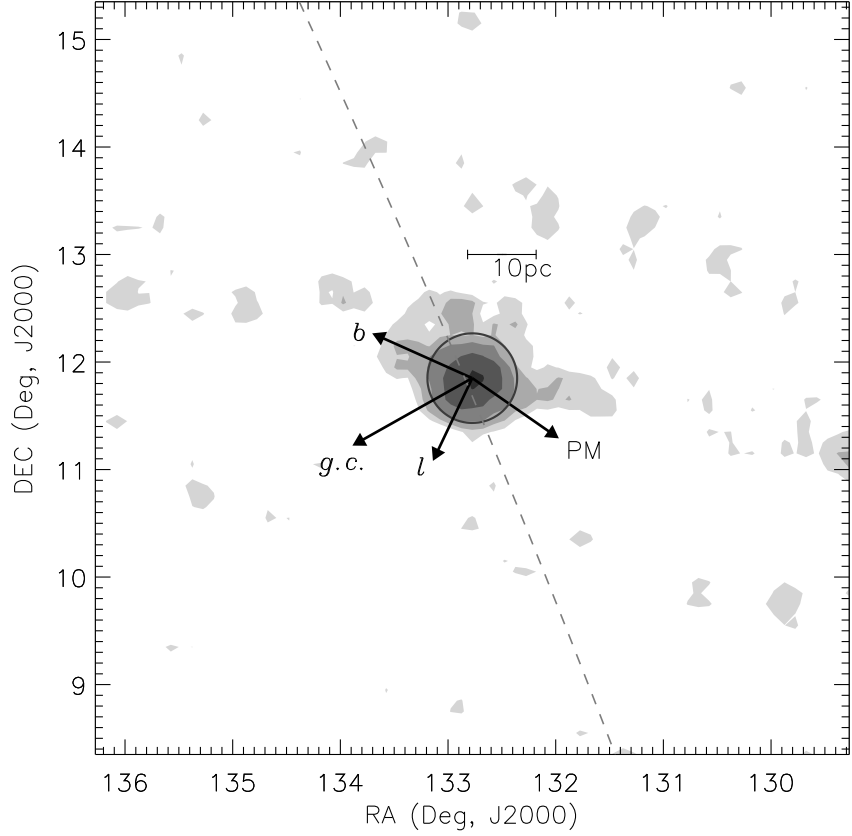}
\caption{The spatial map of high probability cluster members for M67. Darker contours represent increasing exponential density of members, $\alpha$. 
 The contours are defined by the equation $level_j  = MED(\alpha) + (\sigma_\alpha\times2^j)$ as defined in the text. This produces the first contour at 2$\sigma$.
Galactic $(l,b)$ coordinate arrows  and the corrected proper motion vector (PM) from \cite{2008AJ....136..118F} have been added for reference. The arrow labeled $g.c.$ points towards the galactic center. The dark circle denotes the furthest cluster membership in BB03, approximately $0.5^\circ$ in radius. For scale, the bar measures the apparent size of 10pc at M67, a distance of 900pc. The dashed line shows the projection along the orbit which tidal debris would be expected to be scattered.}
\label{map}
\end{figure}

\begin{figure}[]
\centering
\includegraphics[width=3in]{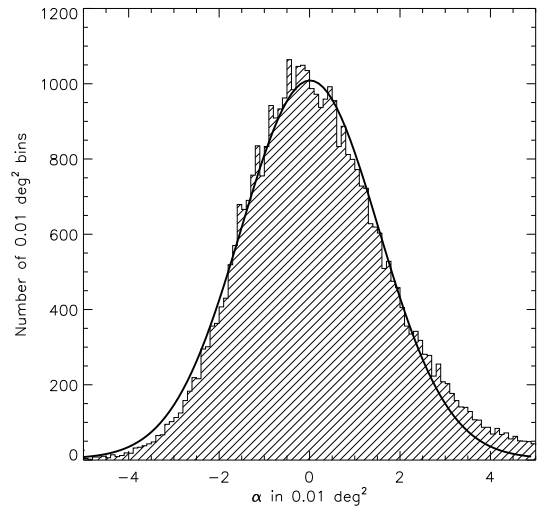}
\caption{The resulting background $\alpha$ counts, in stars per 0.01 deg$^2$.  The entire sample has had a low-order polynomial surface fit subtracted which removes both large scale stellar density gradients (arising from known galactic structure components) and the bias from the response of the field stars to the matched filter.}
\label{alphabkgd}
\end{figure}

\begin{figure}[]
\centering
\includegraphics[width=5in,angle=90]{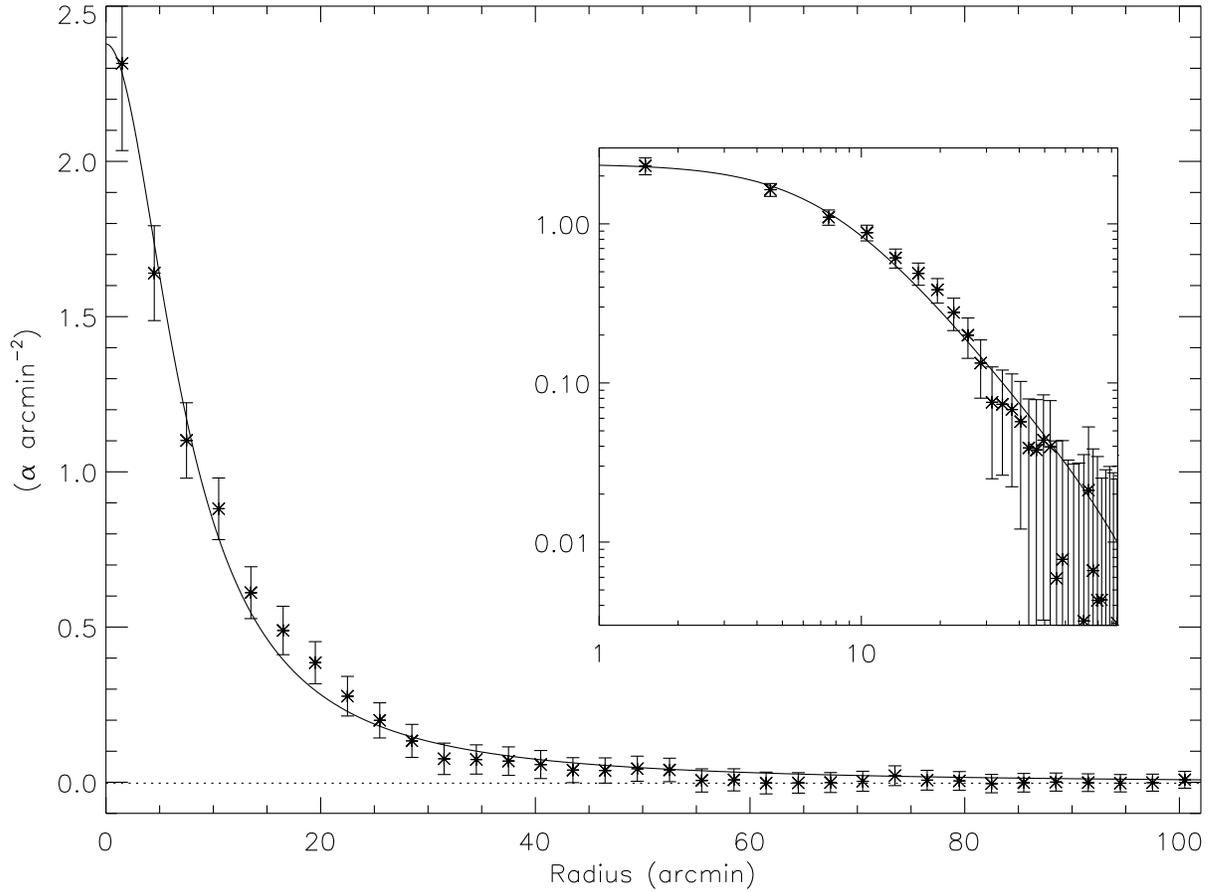}
\caption{The radial density distribution of stars around M67. The vertical axis is the sum of $\alpha$ values  inside each radial bin. The solid line is a King-like profile fit which is described in the text. The dotted line is the fit to the data past $60'$. Error bars are poisson $\sqrt{N}$ errors from the number of stars divided by the area of each radial bin. The insert is the same data shown in log-log space which is often used for cluster density profile studies.
}
\label{surface}
\end{figure}

\begin{figure}[]
\centering
\includegraphics[width=5in]{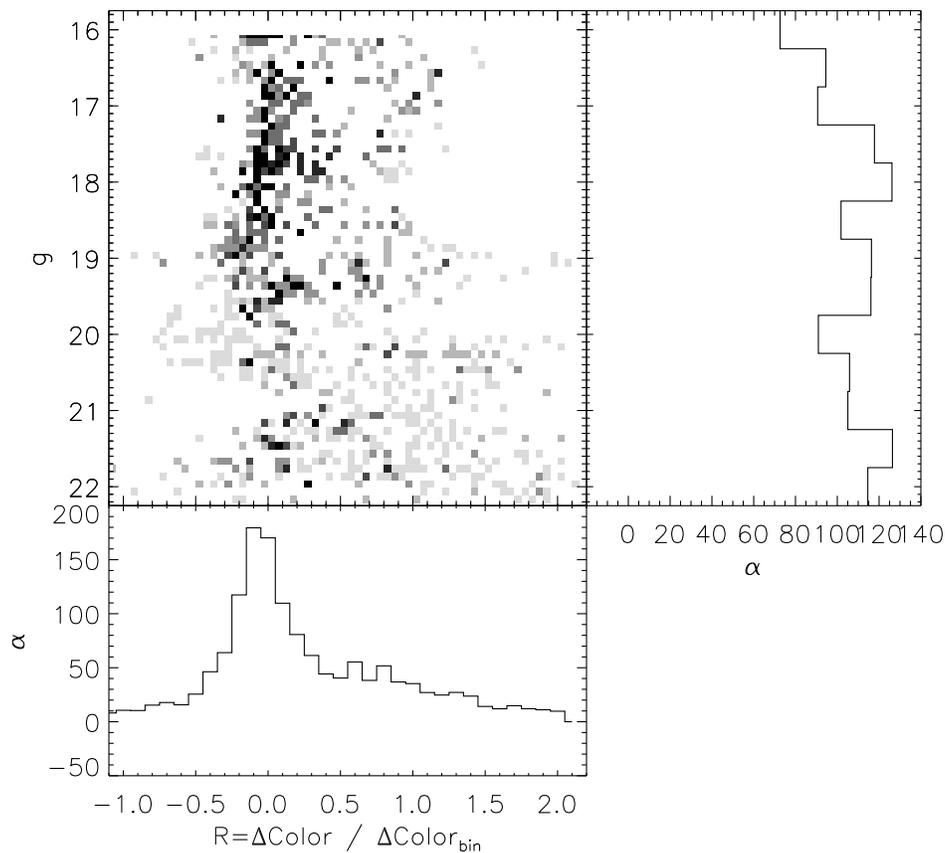}
\caption{The core of M67 (radius $< 0.5^\circ$). A reduced color $(R_{g-i},g)$ CMD is shown in the center panel, where the pixels represent the sum of the $h$ values for every star in each CMD bin for this region. These $h$ values are summed over magnitude space (right) showing the measured luminosity function, and reduced color space (bottom). The bottom panel can be used to estimate the number of stars in the main sequence (centered at $R=0$) and equal mass binary main sequence ($R=1$).}
\label{core}
\end{figure}

\begin{figure}[]
\centering
\includegraphics[width=5in]{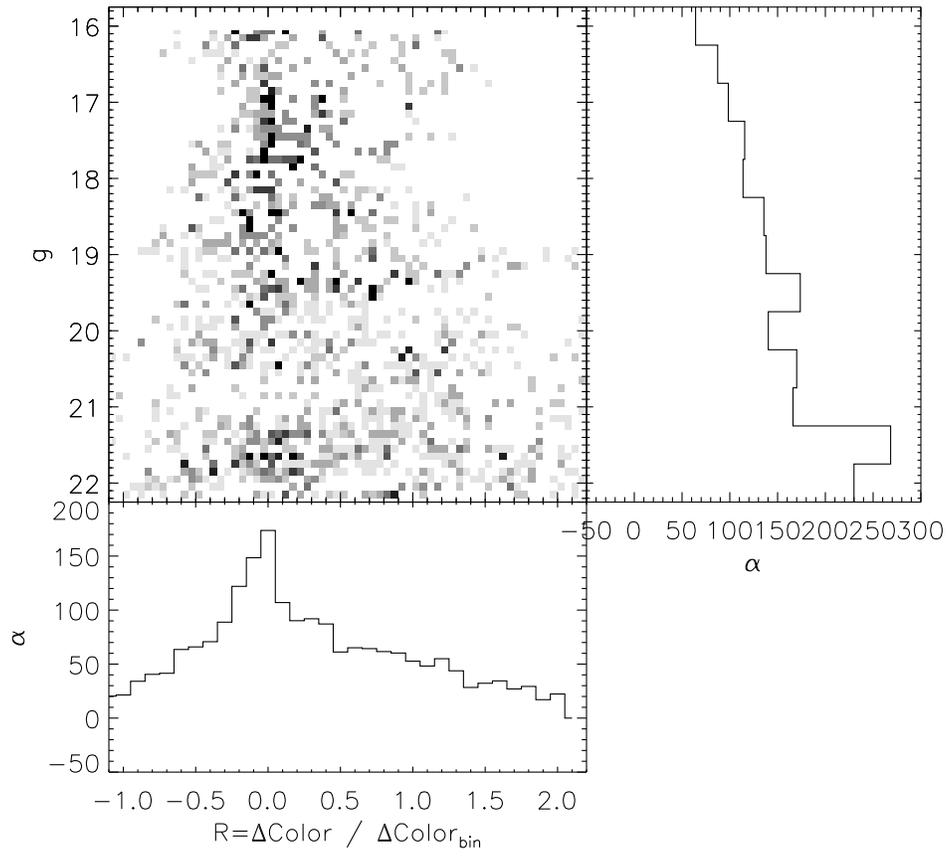}
\caption{Same as Figure \ref{core} for the halo around M67 ($0.5^\circ<$ radius $<1.2^\circ$).}
\label{halo}
\end{figure}

\begin{figure}[]
\centering
\includegraphics[width=5in]{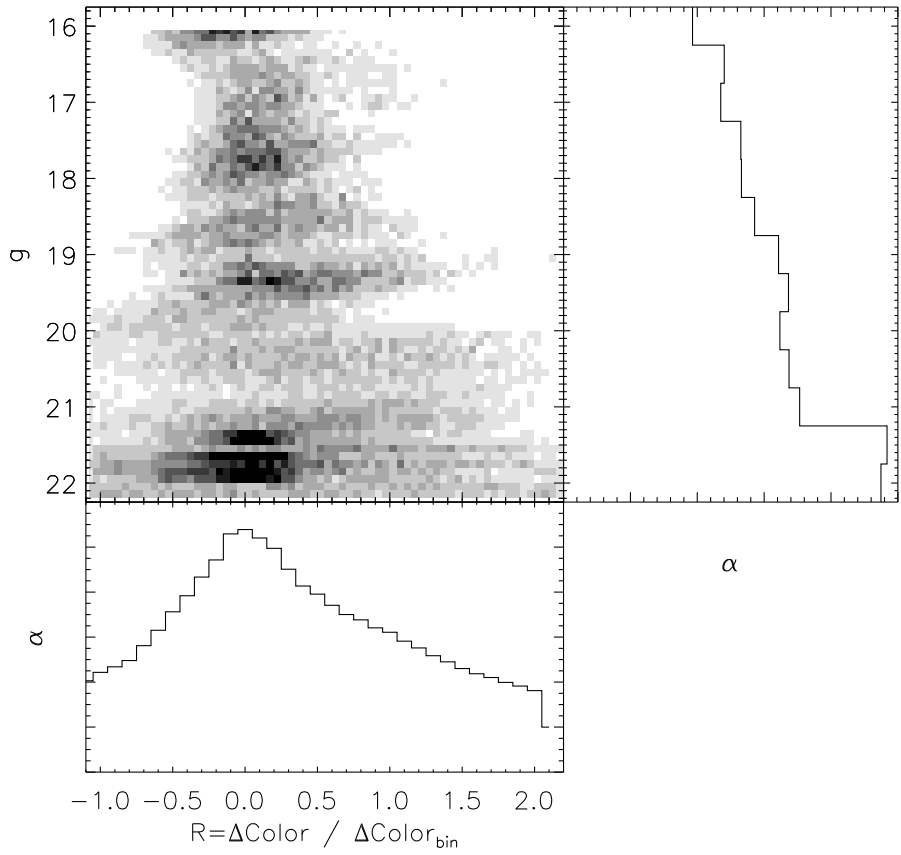}
\caption{Same as Figure \ref{core} for the background around M67 ($1.5^\circ<$ radius $<8^\circ$).}
\label{bkgd}
\end{figure}

\begin{figure}[]
\centering
\includegraphics[width=5in,angle=90]{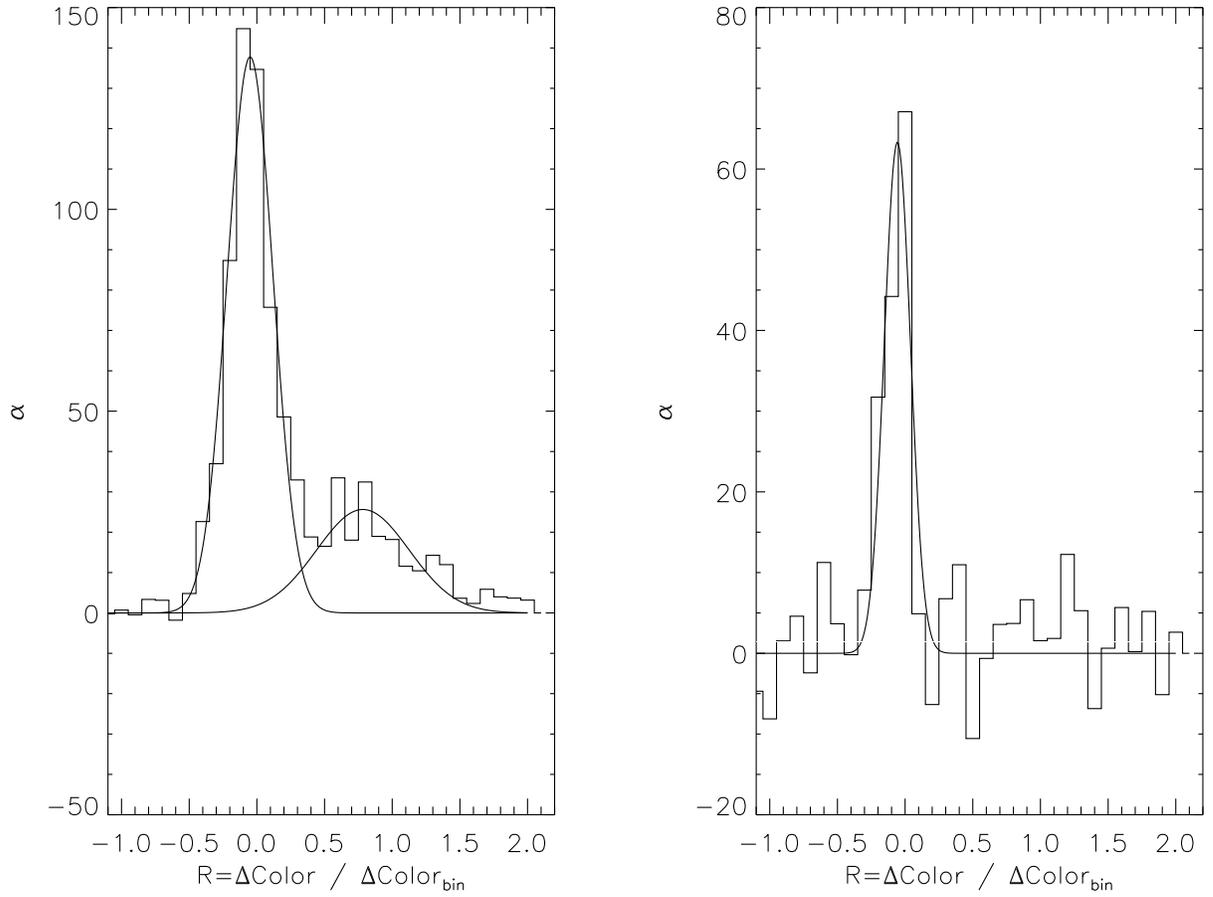}
\caption{The relative binary fraction in the core (left) and halo (right). The single-star main-sequence is centered at $R=0$, while the equal-mass binary  main-sequence is at $R=1$. Note the lower recovery of binary stars in the halo.}
\label{bf}
\end{figure}

\begin{figure}[]
\centering
\includegraphics[width=5in,angle=90]{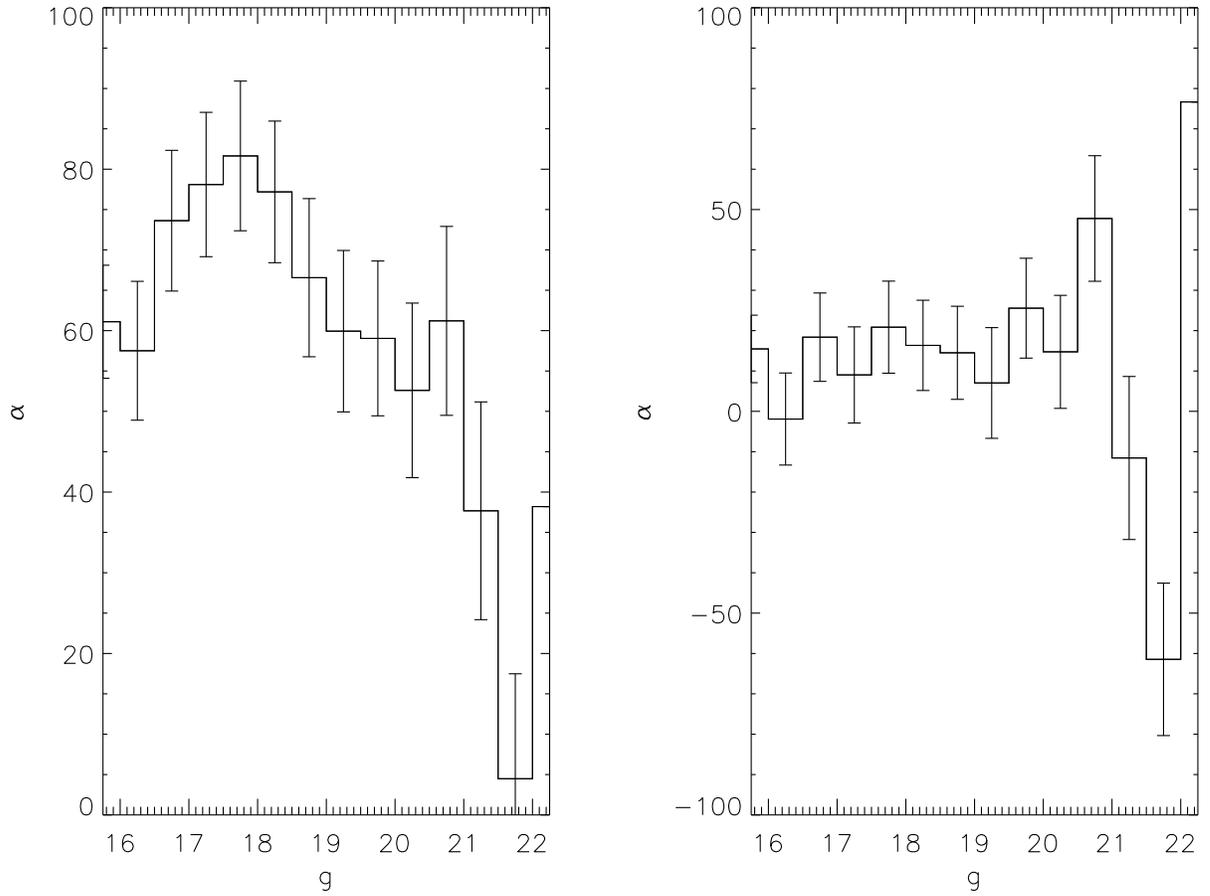}
\caption{The relative luminosity functions in the core (left) and halo (right). The core is deficient of lower-mass stars compared to the halo due to ongoing energy equipartition within the cluster.}
\label{lf}
\end{figure}

\begin{figure}[]
\centering
\includegraphics[width=5in]{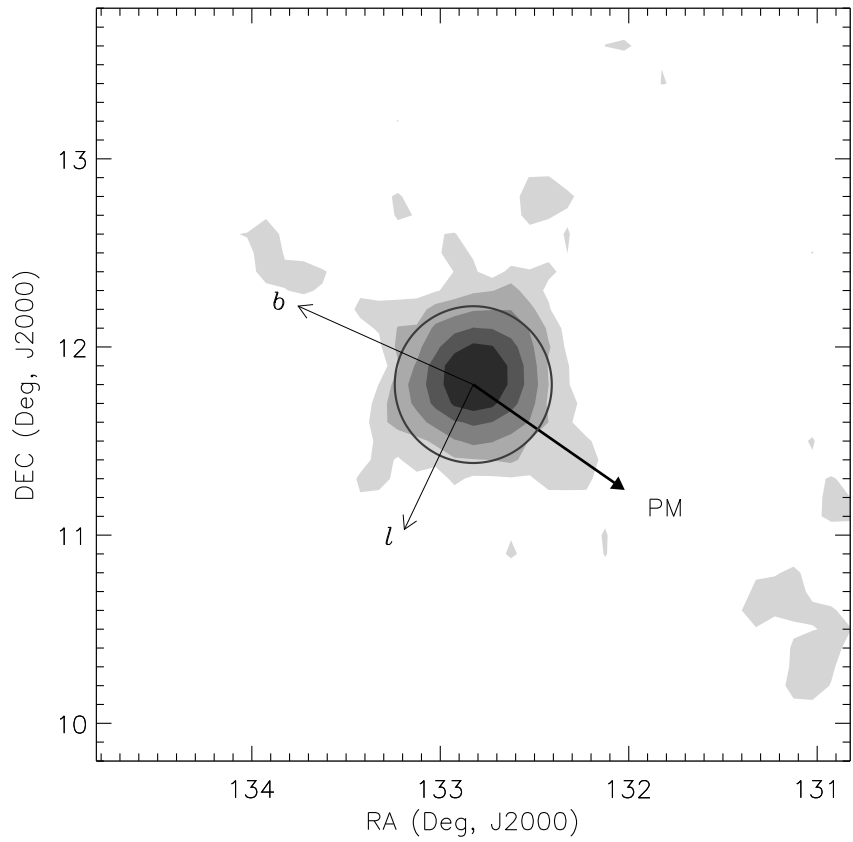}
\caption{The spatial map of high probability cluster members for M67 using 2MASS. This figure has the same contours and labeling as Figure \ref{map}.}
\label{2mass1}
\end{figure}

\begin{figure}[]
\centering
\includegraphics[width=5in,angle=90]{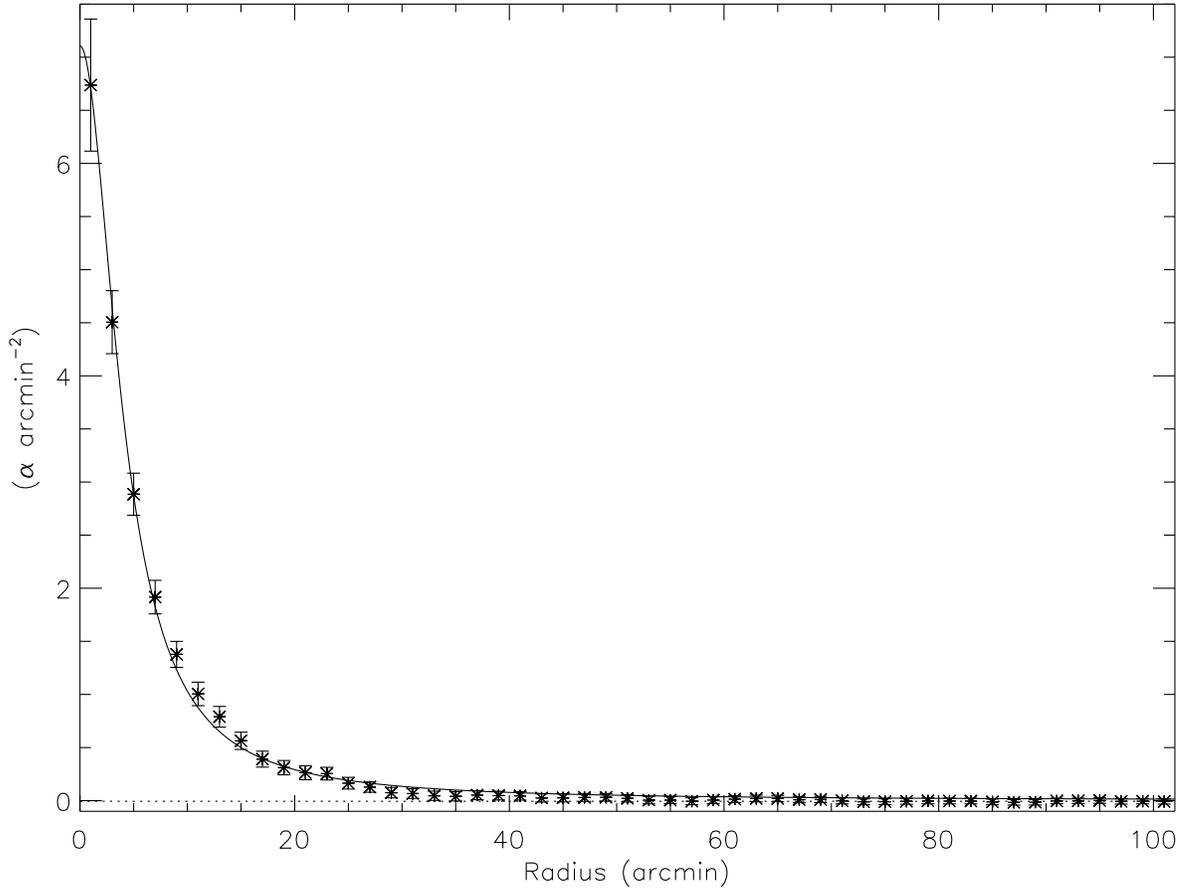}
\caption{The radial distribution of stars around M67 with 2MASS. This figure has the same fit and error information as Figure \ref{surface}.}
\label{2mass2}
\end{figure}

 \citet{bergond01} have suggested that a general alignment between the proper motion and the elongation of an open cluster is not uncommon. They also demonstrate that the major axis in the core of a cluster may preferentially point toward the galactic center.  The second highest contour in Figure \ref{map} in the core of M67 is slightly elongated along the direction of the galactic center. Our newly discovered tidal tails also emanate almost perpendicular to this as seen in  \citet{bergond01}, and our tidal features seem to fit all expectations of an open cluster being pulled apart.

  \citet{cc94} project the orbit of M67 back in time to approximately its birth. Their calculations show that M67 has made approximately 17 passes through the galactic plane ($Z=0$), and that the current position of M67 is near the vertical apex of its orbit ($Z=0.41$Kpc). 
It is generally thought that the large distance above the galactic plane at which M67 spends much of the time has helped prevent tidal effects from completely destroying the cluster, as is the case with NGC 6791.

\begin{figure}[]
\centering
\includegraphics[width=5in]{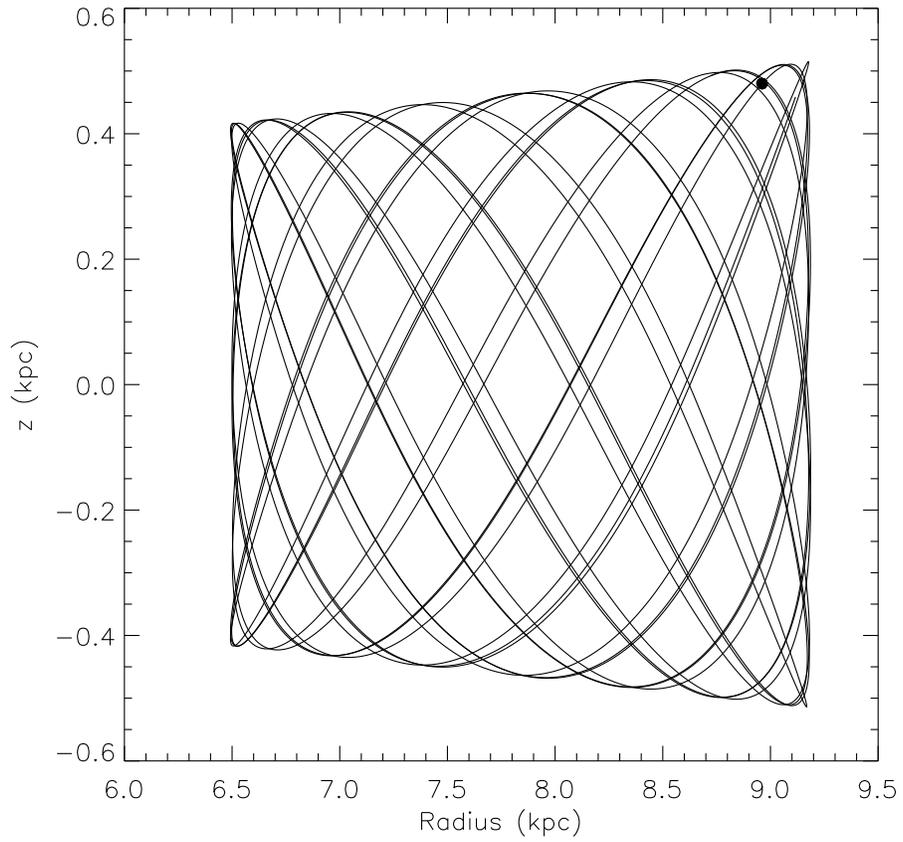}
\caption{The meridional plane orbit of M67, projected 1Gyr in the past and future. The present location of M67 is denoted by the dark circle.}
\label{rz_orbit}
\end{figure}

Using modern determinations of the proper motions, radial velocity, and distance to M67 from \citet{2008AJ....136..118F}, we calculated the cluster's full galactic orbit. We followed the same procedure as \citet{cc94}, which employed a three component galactic potential, and used the \citet{1987AJ.....93..864J} method of computing the space velocities. Figure \ref{rz_orbit} shows  the meridional projection for the orbit of M67, calculating 5\e{8} years into the past and future. This is in excellent agreement with \citet{cc94} and earlier estimations.

Typically long streams of tidal debris, such as those trailing/leading Pal 5, are assumed to lie along the orbital path of the cluster\citep{montuori07}. However this assumption has recently come under scrutiny for Pal 5 \citep{2009pal5}.  \citet{bergond01} also show examples of tidal debris around open clusters which is not grouped into dynamically cold streams. In order to determine the projected direction tidal debris tails would be expected to lie along for M67, we
computed star paths in a reference frame moving with the average
orbital velocity of the cluster. Simply put, stars that are leading or
lagging the main body of the cluster in its orbit around the galaxy do
not diffuse to completely ring the galaxy because they mostly share
the orbital motion of the cluster. Rather they will occupy a more
modest range of azimuthal angles in the galaxy, with smaller ranges
corresponding to smaller orbital eccentricity. 
M67 moves both in the radial and vertical directions (as shown in Fig. \ref{rz_orbit}).
For
an open cluster orbiting with vertical excursions from the disk, the
orbit will not be closed, and the cluster (and dynamically cold stars
that have escaped the cluster) will eventually sample different parts
of a boxy volume moving with the average orbital velocity of the
cluster. To visualize this three-dimensional dynamical structure, we
subtracted the average angular velocity of the cluster multiplied by the time
since the present day from the azimuth angle.
The projection of this motion onto our line of sight towards M67 looks nearly straight, as shown in Fig. \ref{map}. Any kinks or sharp changes in the projected co-moving deviation are well outside our SDSS field. 

Since the most elongated feature of M67's halo seen in Fig. \ref{map} is not aligned with the projected co-moving deviation, or the projected orbital path of M67, we believe the asymmetric tidal feature's origins seen with SDSS cannot be singularly attributed to the galactic tidal field. However a separate smoking gun, such as a nearby molecular cloud or cluster, is not readily apparent to us. It is more probable that with such small numbers of stars found outside the core, tidal shocking from passing through the galactic disk produces these types of weakly elongated features in open clusters, rather than the dramatic tails as seen in Pal 5. This can only be verified, however, using a detailed survey of the kinematics of the cluster members on a large spatial scale to compare the core and halo dynamics, along with future N-body modeling of cluster disruption. 

The optimal contrast filtering we have employed here will be of great use in the next generation of wide field surveys. Programs such as LSST \citep{lsst} and PanSTARRS \citep{panstarrs} will map the sky with never before seen levels of photometric precision and depth, with a spatial coverage far exceeding even SDSS. The most recent SDSS public release \citep[DR7]{dr7} includes the Sloan Extension for Galactic Understanding and Exploration \citep{segue} data, which includes photometry in and around the galactic plane. This new data set includes well over a hundred open clusters, with a great range in ages, masses, and surely dynamical states. We anticipate the matched filtering technique will help detect many new open and globular cluster features, as well as continue to find other faint substructure in the galactic halo.

\acknowledgements 
The authors gratefully acknowledge the support of this work by the National Science Foundation grant AST 0507785, awarded to ELS \& M. Bolte. JRAD would like to thank Dr. Suzanne L. Hawley for her illuminating discussions on stellar populations and cluster modeling. JRAD and ELS thank Dr. Kathryn Johnston for her help with galactic orbit calculations.


\clearpage

\clearpage

\clearpage

\clearpage

\clearpage

\clearpage

\clearpage

\clearpage

\clearpage

\clearpage

\clearpage

\clearpage

\clearpage

\clearpage

\clearpage

\clearpage

\clearpage

\clearpage

\clearpage

\clearpage

\end{document}